\documentclass{article}

\usepackage{amsmath,amsfonts,amsthm}
\usepackage{graphicx}
\usepackage{fixmath}
\usepackage{amssymb,latexsym}
\usepackage{mathrsfs,amsbsy}
\usepackage{dsfont}
\usepackage{algorithm}
\usepackage{eucal}
\usepackage{colortbl}
\usepackage{multirow}
\usepackage[end]{algpseudocode}
\usepackage{float}
\graphicspath{{./figures/}}
\usepackage{booktabs}
\usepackage{caption}
\usepackage{subfigure}
\usepackage{microtype}

\theoremstyle{definition}

\theoremstyle{remark}

\numberwithin{equation}{section}
\numberwithin{subsection}{section}

\title{
A Novel Deformation Method for Higher Order Mesh Generation
}
\author{Zicong Zhou\\
University of Texas at Arlington\\
zicong.zhou@mavs.uta.edu
\and
Xi Chen\\
xi.chen@mavs.uta.edu 
\and 
Guojun Liao\\
University of Texas at Arlington\\
liao@uta.edu
}

\begin{document}
\maketitle    
\begin{abstract}
The development of higher order finite elements methods has become an active research area \cite{Solin}. The $deformation$ $method$ for mesh generation has achieved a prescribed positive Jacobian determinant constraint \cite{Ands} \cite{Pan} and it has been a useful method for mesh adaptation \cite{Liu} \cite{Liseikin} \cite{turek}. In this work, the $deformation$ $method$ is combined with local refinement to generate higher order meshes. 2D and 3D numerical examples of element order $p=3$ are shown to demonstrate the effectiveness of our new algorithm.
\end{abstract}

\section{Introduction}
The study of diffeomorphisms is an important research area of differential geometry. In 1990s, J. Moser and B. Dacorogna studied the existence and construction of diffeomorphisms under a positive Jacobian determinant constraint on a domain in $\mathbb R^n$ \cite{Dac:Mos}. The deformation approach in their study has been extended to the adaptive mesh generation problem by our group and collaborators \cite{Liu}. The higher order finite elements methods has become one of the attended fields. In this paper, we further extend the $deformation$ $method$ to higher order mesh. 

In \cite{Pan} and \cite{Liu}, the initial version of the method are based on Poisson equations which can only work on fixed domains. In \cite{Cai} and  \cite{Dion}, a $\bf{divergence-curl}$ system is formulated which solves the mentioned positive prescribe Jacobian problem in variable domains. 

In this paper, we propose a new algorithm based on the $deformation$ $method$ to generate higher order elements on a domain $\mathrm{\Omega}$ by the following steps: (1) Form a coarse (linear) mesh on the domain $\mathrm{\Omega}$; (2) Subdivide the coarse mesh to an intermediate mesh; (3) Deform the intermediate (linear) mesh by moving the boundary nodes of the intermediate mesh to the boundary of $\mathrm{\Omega}$, thus a new boundary conforming mesh is generated; (4) Use the new locations of the nodes to generate a higher order mesh by interpolation. 

In section 2, the deformation method based on $\bf{divergence-curl}$ system is reviewed. In section 3 more details for step 2 to step 4 as well as numerical examples for $p=3$ are provided. 

\section{The $deformation$ $method$ for variable domains}
Let $\mathrm{\Omega}_t \subset \mathbb R^n$, $n=2$, or $3$ and $0\leq{t}\leq{T}$, be a moving (includes fixed) domain and $\pmb{v}(\pmb{x},t)$ be the velocity field on $\partial\mathrm{\Omega_t}$, where $\pmb{v}(\pmb{x},t)\cdot{\pmb{\mathrm{n}}}=0$ on any part of $\partial\mathrm{\Omega}_t$ with slippery-wall boundary conditions and $\pmb{\mathrm{n}}$ is the outward normal vector of $\partial\mathrm{\Omega}_t$. Given scalar function $f(\pmb{x},t)>0 \in C^1(\pmb{x},t)$ on the domain $\mathrm{\Omega}_t \times [0,T]$, such that

	\begin{equation}\label{eq:plm1}
	f(\pmb{x},0)=1,
	\end{equation}
	\begin{equation}
	\int_{\mathrm{\Omega}_t} \dfrac{1}{f(\pmb{x},t)}d\pmb{x} = |\mathrm{\Omega}_0|.
	\end{equation}
We look for a diffeomorphism 

	\begin{equation} 
	\pmb{\phi}(\pmb{\xi},t):\mathrm{\Omega}_0\rightarrow\mathrm{\Omega}_t
	\end{equation}
such that $\forall t \in [0,T]$

	\begin{equation}\label{eq:plm2}
	J(\pmb{\phi}(\pmb{\xi},t)) = f(\pmb{\phi}(\pmb{\xi},t),t)
	\end{equation}
where $J(\pmb{\phi}(\pmb{\xi},t)) =\text{det}\bigtriangledown(\pmb{\phi}(\pmb{\xi},t)).$ 
\begin{figure}[H]
		\begin{center}
		{\includegraphics[width=9cm,height=4.5cm]{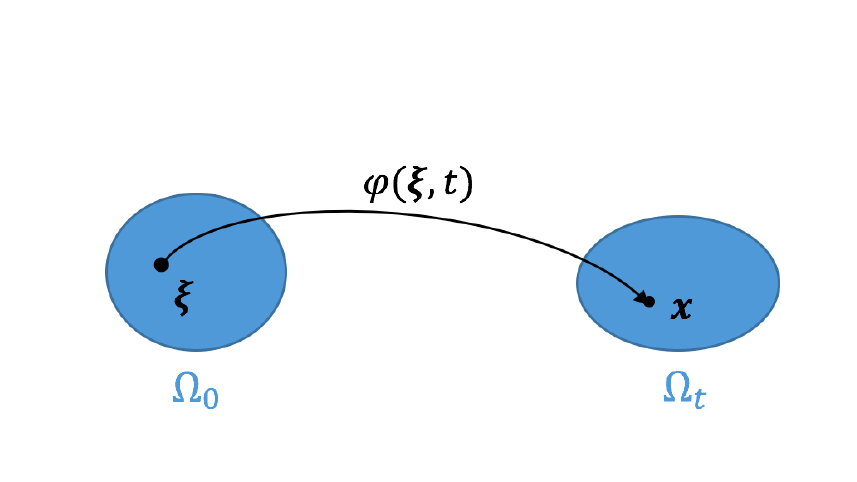}}
		\end{center}
		\caption{Problem illustration: $\pmb{\varphi}$ maps $\pmb{\xi}$ to $\pmb{x}$}
\end{figure}
	
In \cite{Cai} and \cite{Dion}, it is shown that such diffeomorphism $\pmb{\phi}$ can be constructed by solving the following differential equations: 
	\begin{itemize}
	\item determine $\pmb{u}(\pmb{x},t)$ on $\mathrm{\Omega}_t$ by solving
	\begin{equation}\label{eq:sol1}
	\left\{
		\begin{aligned}
		\text{div } \pmb{u}(\pmb{x},t)& = -\frac{\partial}{\partial t}(\dfrac{1}{f(\pmb{x},t)}) \\
		\text{curl } \pmb{u}(\pmb{x},t)& = 0\\
		\pmb{u}(\pmb{x},t)& = \dfrac{\pmb{v}(\pmb{x},t)}{f(\pmb{x},t)} \text{, on } \partial\mathrm{\Omega}_t
		\end{aligned}\right.
	\end{equation}
	\item determine $\pmb{\phi}(\pmb{\xi},t)$ on $\mathrm{\Omega}_0$ by solving	
		\begin{equation}\label{eq:sol2}
		\left\{
			\begin{aligned}
			\frac{\partial \pmb{\phi}(\pmb{\xi},t)}{\partial t}& = f(\pmb{\phi}(\pmb{\xi},t),t) \pmb{u}(\pmb{\phi}(\pmb{\xi},t),t), \\
			\pmb{\phi}(\pmb{\xi},0)& = \pmb{\xi}
			\end{aligned}\right.
		\end{equation}
	\end{itemize}
	
Based on this two-step procedure we may construct a diffeomorphism to the problem of $(\ref{eq:plm1})$ to $(\ref{eq:plm2})$. The related theoretical derivations can be traced back from \cite{XiChen}. The constraint $(\ref{eq:plm1})$ simply indicates that the construction of such diffeomorphism is started and deformed from $\pmb{\phi}(\pmb{\xi},0)=\pmb{id}(\pmb{\xi})$.  Moreover, from the perspective of numerical computations, all the simulations are built on discretized domains, therefore it is naturally feasible to revise such mesh-like configuration of simulations into a method of mesh generation and it is named as the $deformation$ $method$ for mesh generation.

It turns out, the most effective numerical method to solve the $\bf{divergence}$ $\bf{-curl}$ system $(\ref{eq:sol1})$, in terms of accuracy, flexibility and compatibility, is the Least-Squares Finite-Element Method (LSFEM) \cite{Cai} \cite{Dion} \cite{Jiang}. As for $(\ref{eq:sol2})$, any standard numerical ODE methods can be used. So, for simplicity we choose the $Explicit$-$Euler$ method here. We had proposed the following Algorithm for the $deformation$ $method$ on moving domains.

\begin{algorithm}[H]
	{
	\caption{: $deformation$ $method$ }
	\hrule
	\begin{itemize}
		\item Step 1: $\mathbf{Initialization}$. Start from $t=0$ to $T=1$ and define $\pmb{\phi}(\pmb{\xi},0)= \pmb{\xi}$, $f(\pmb{x},0)=1$
		\item Step 2: $\bf{divergence-curl}$.
		Compute  $-\frac{\partial}{\partial t}(\dfrac{1}{f(\pmb{x},t)})$. 
		Solve (2.5) by LSFEM to get $\pmb{u}(\pmb{x},t)$
		with Dirichlet condition on moving boundary and Neumann condition on slippery wall boundary
		
		\item Step 3:  $\mathbf{ODE}$. Update $\pmb{\phi}(\pmb{\xi},t)$ by (2.6) using $Explicit$-$Euler$ method, from $\mathrm{\Omega}_t$ to $\mathrm{\Omega}_{t+dt}$.
		\item Step 4: Update $t=t+dt$, then go to $\mathbf{Step}$ $\mathbf{2}$ until $t=1=T$
	\end{itemize}
	}
\end{algorithm}

\begin{figure}[H]
	\caption{Deforming a rectangle onto a quarter disk of $radius=1.25$, which center at the origin and locate on the first quadrant. The left and bottom sides are slippery-wall condition. The boxed numbers are the labels of each elements.}
	\begin{center}
	\subfigure[t=0]{\includegraphics[width=6cm,height=6cm]{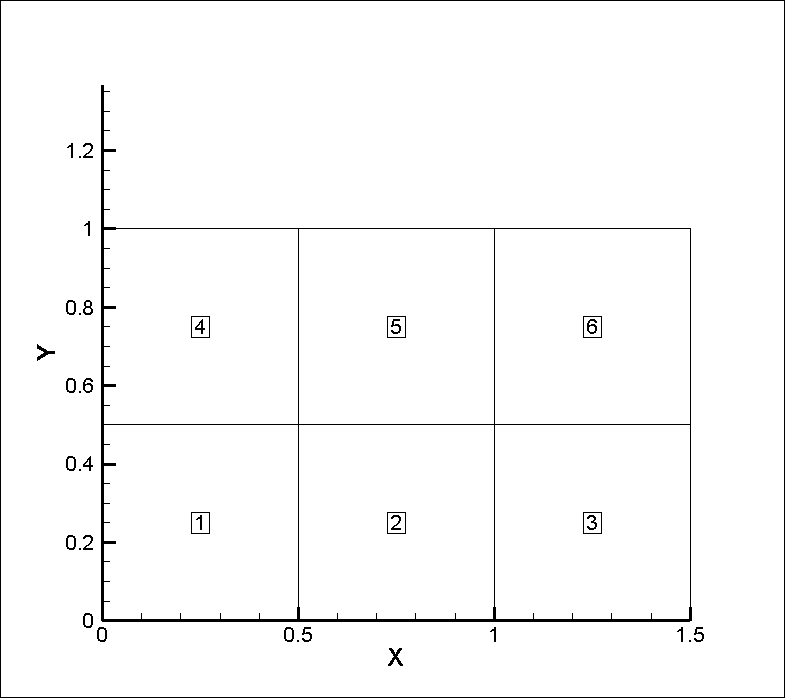}}
	\subfigure[t=0.3]{\includegraphics[width=6cm,height=6cm]{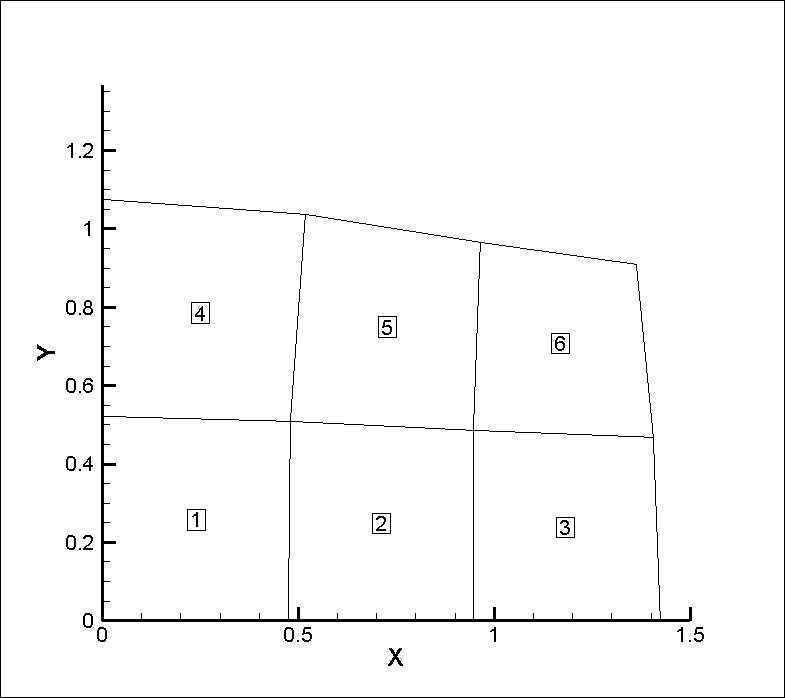}}
	\subfigure[t=0.6]{\includegraphics[width=6cm,height=6cm]{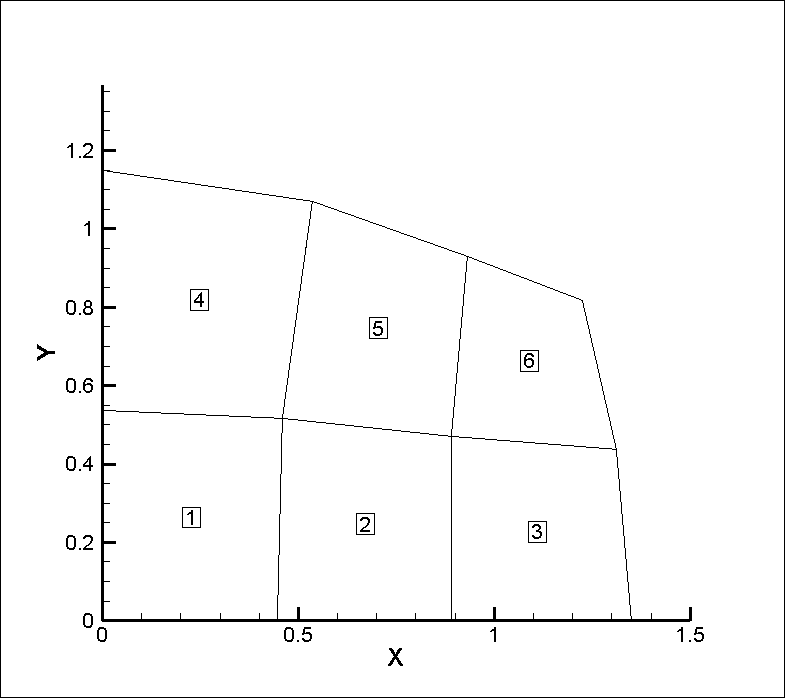}}
	\subfigure[t=1]{\includegraphics[width=6cm,height=6cm]{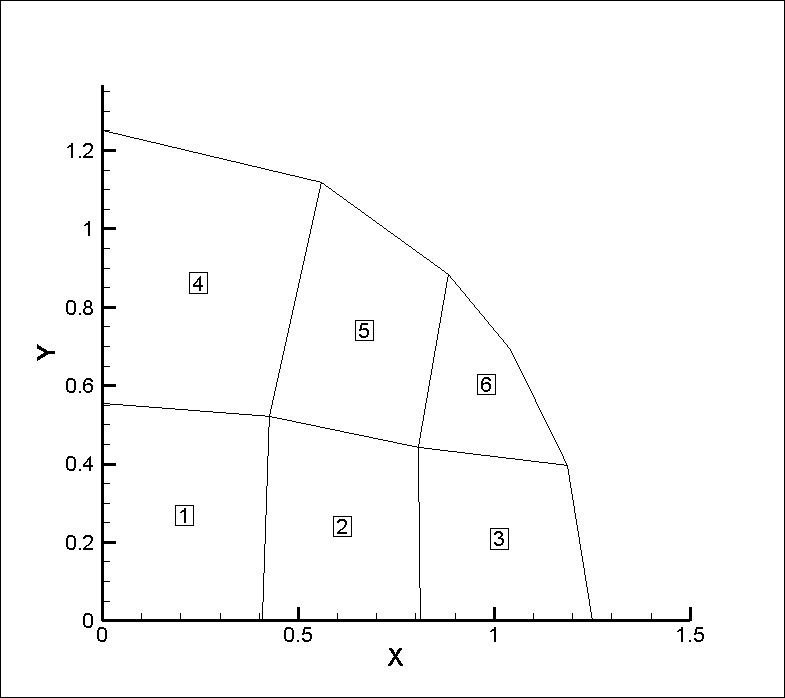}}
	\end{center}
\end{figure}
\begin{figure}[H]
	\caption{Deforming a brick onto a quarter upper semi-ball that centers at the origin and locate on positive side of $\mathbb R^3$. Those 3 faces on the back are slippery-wall condition.}
	\begin{center}
	\subfigure[t=0]{\includegraphics[width=6cm,height=6cm]{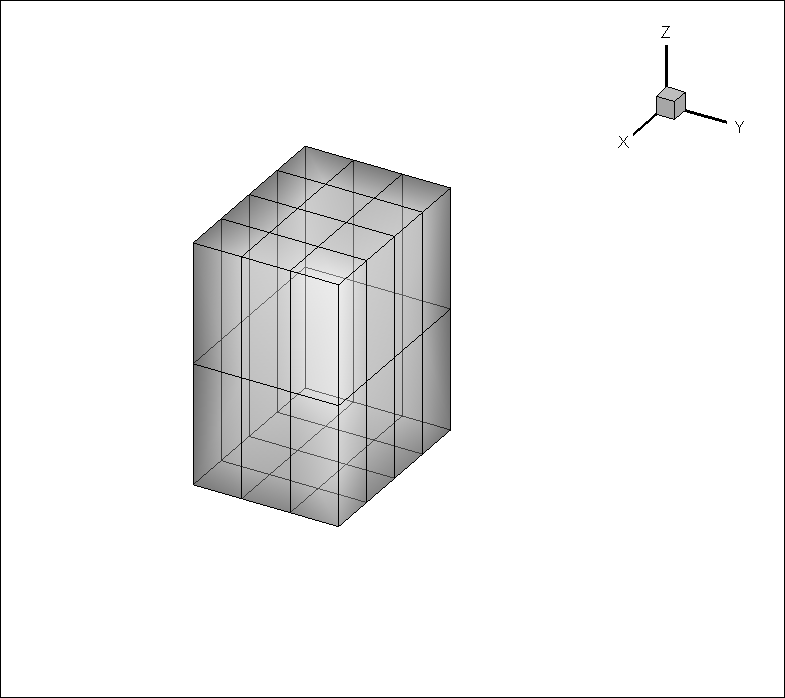}}
	\subfigure[t=0.3]{\includegraphics[width=6cm,height=6cm]{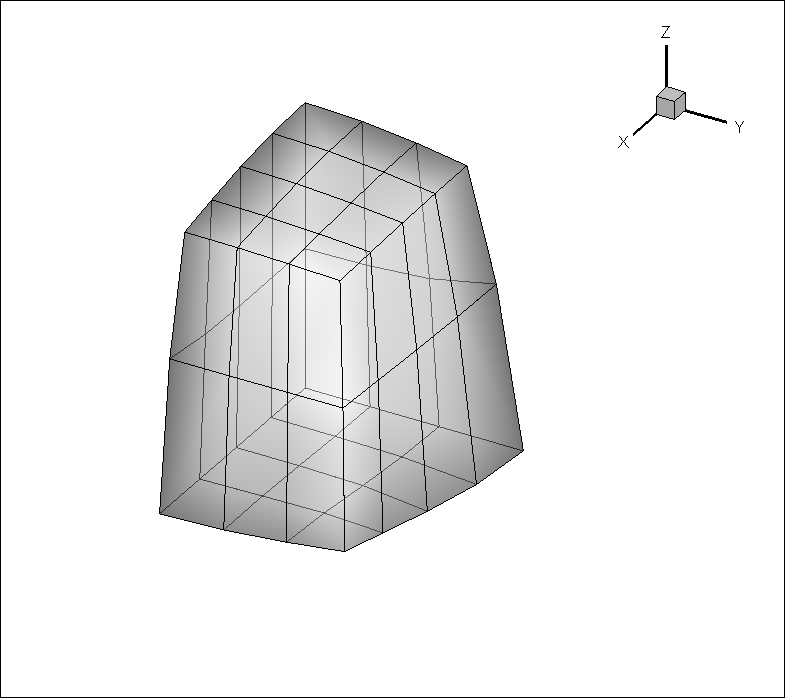}}
	\subfigure[t=0.6]{\includegraphics[width=6cm,height=6cm]{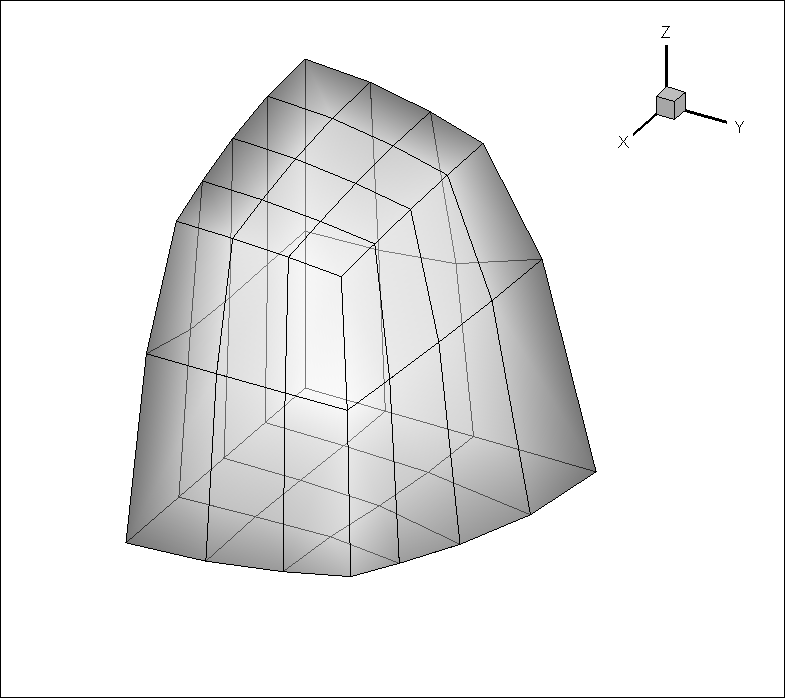}}
	\subfigure[t=1]{\includegraphics[width=6cm,height=6cm]{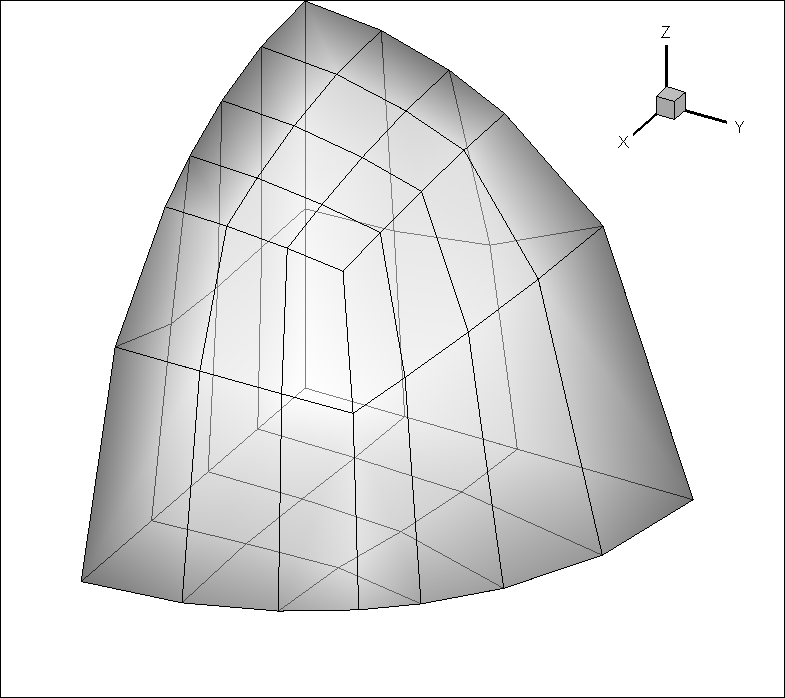}}
	\end{center}
\end{figure}

A few more examples demonstrate flexible prescribed Jacobians both 2D \& 3D:
\begin{figure}[H]
	\caption{The prescribed Jacobian $f(\pmb{x},t)$ is defined by an image of a basketball.}
	\begin{center}
	\subfigure[t=0]{\includegraphics[width=6cm,height=5.75cm]{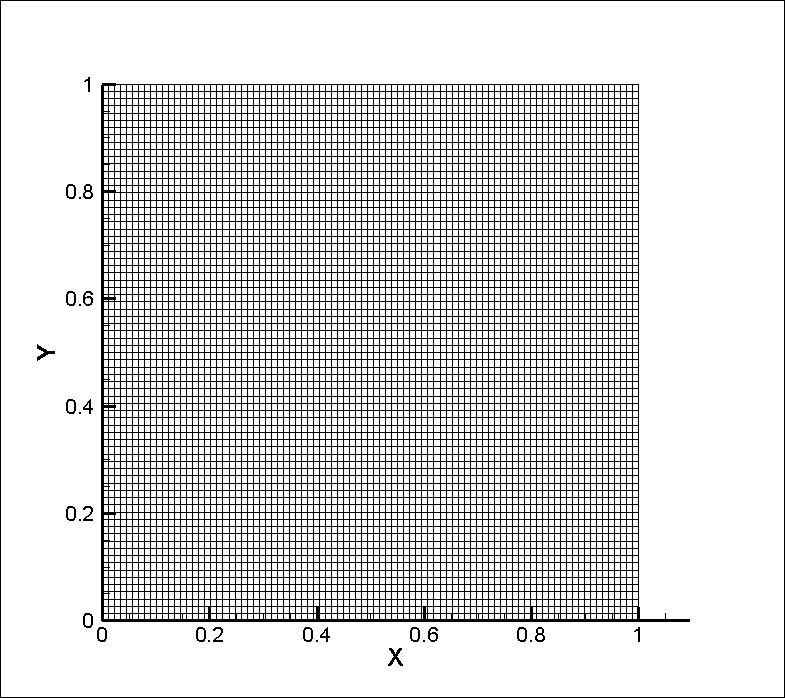}}
	\subfigure[t=0.5]{\includegraphics[width=6cm,height=5.75cm]{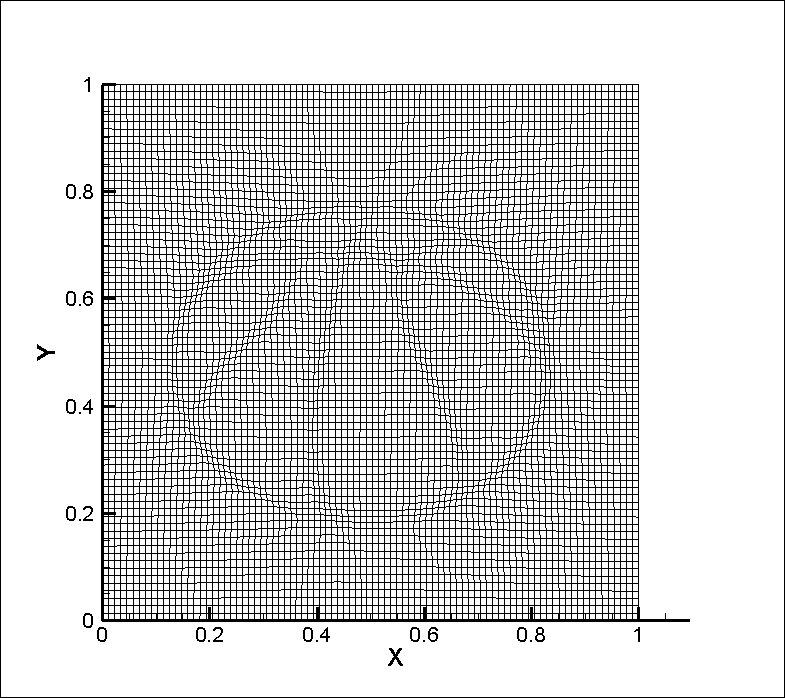}}
	\subfigure[t=1]{\includegraphics[width=10cm,height=9.25cm]{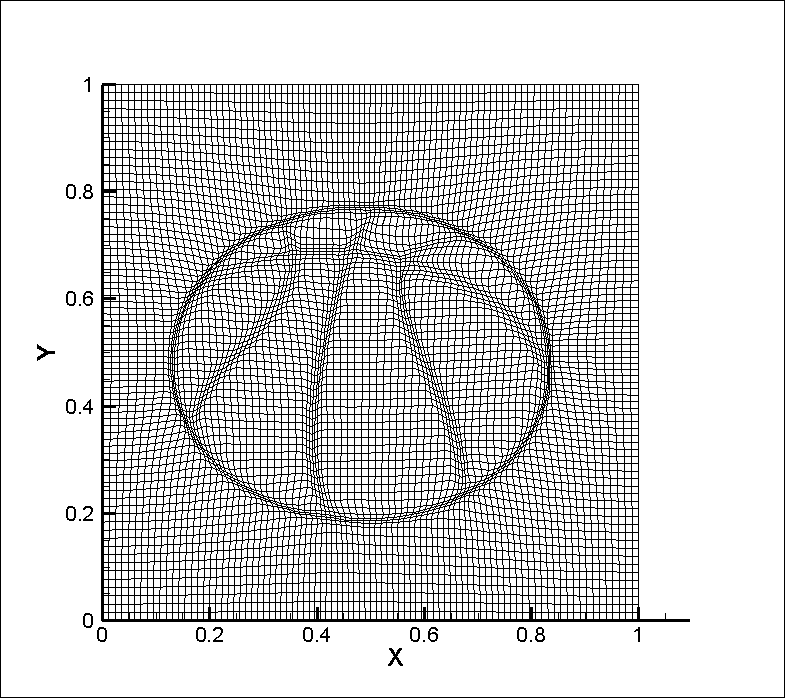}}
	\end{center}
\end{figure}
\begin{figure}[H]
	\caption{The prescribed Jacobian $f(\pmb{x},t)$ describes an interface of an ellipse for the interior of $\mathrm{\Omega}_t$ while the boundaries were deformed to another ellipse.}
	\begin{center}
	\subfigure[t=0]{\includegraphics[width=6cm,height=6cm]{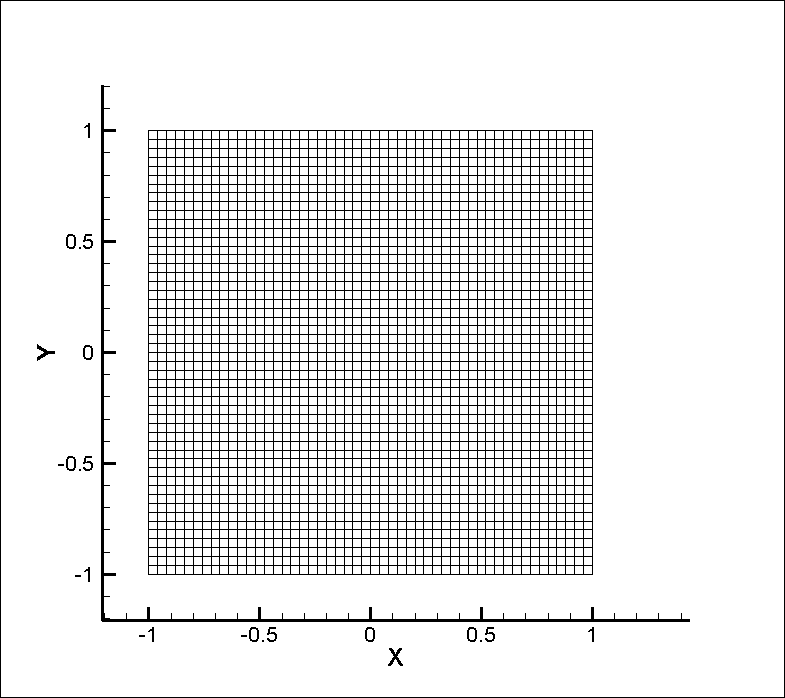}}
	\subfigure[t=0.5]{\includegraphics[width=6cm,height=6cm]{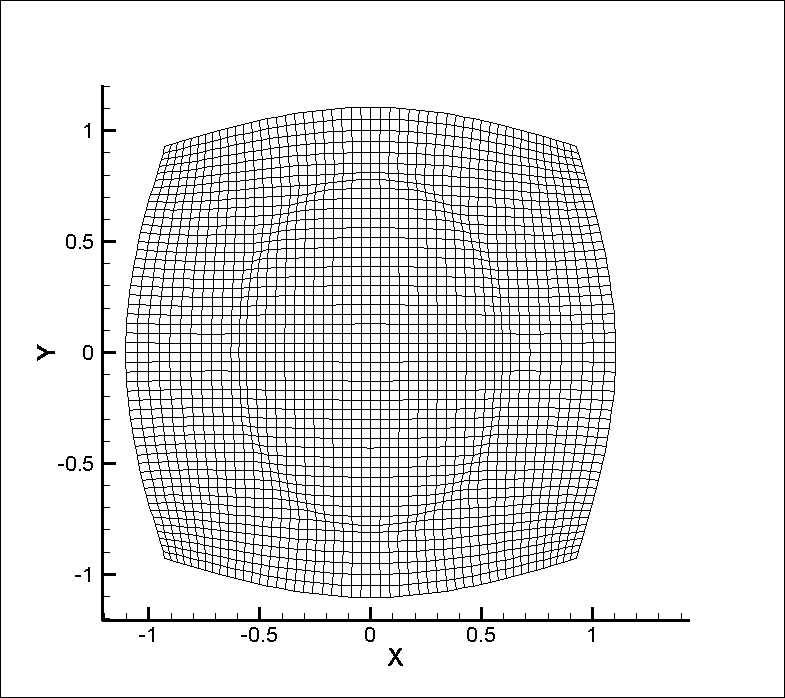}}
	\subfigure[t=1]{\includegraphics[width=9cm,height=9cm]{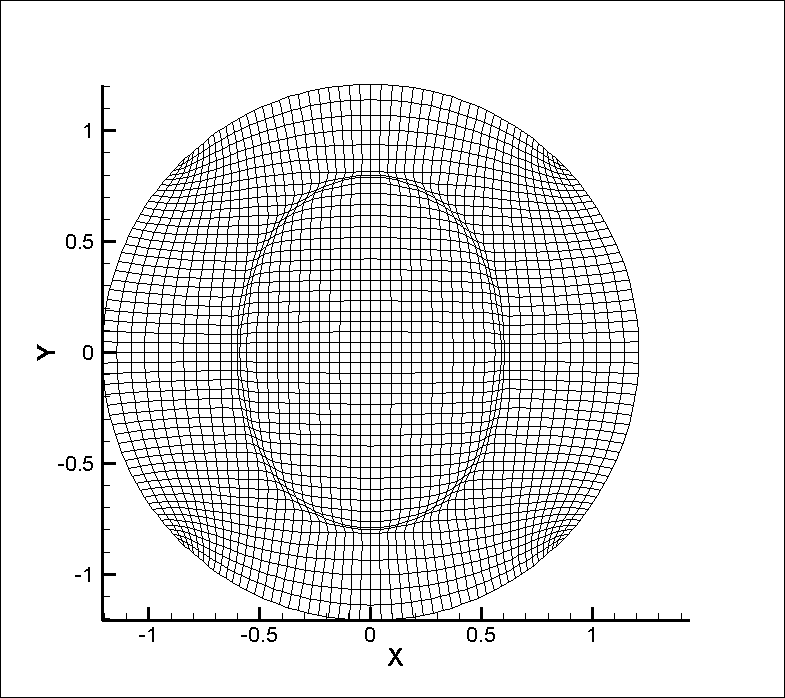}}
	\end{center}	
\end{figure}
\begin{figure}[H]
	\caption{Pushing a cube to sit on an invisible ball while emphasizing the top face with an interface of another ball}
	\begin{center}
	\subfigure[t=0]{\includegraphics[width=6cm,height=5.5cm]{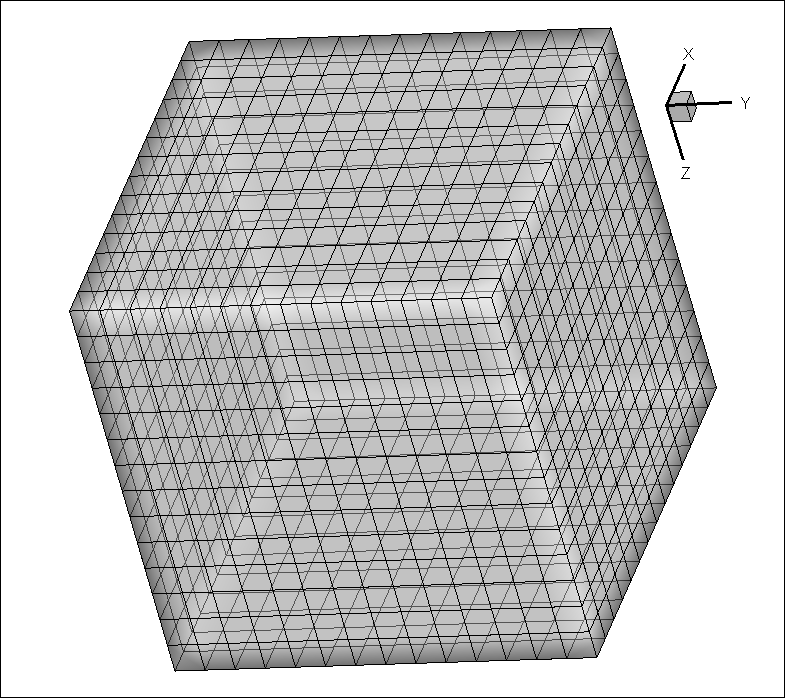}}
	\subfigure[t=0.5]{\includegraphics[width=6cm,height=5.5cm]{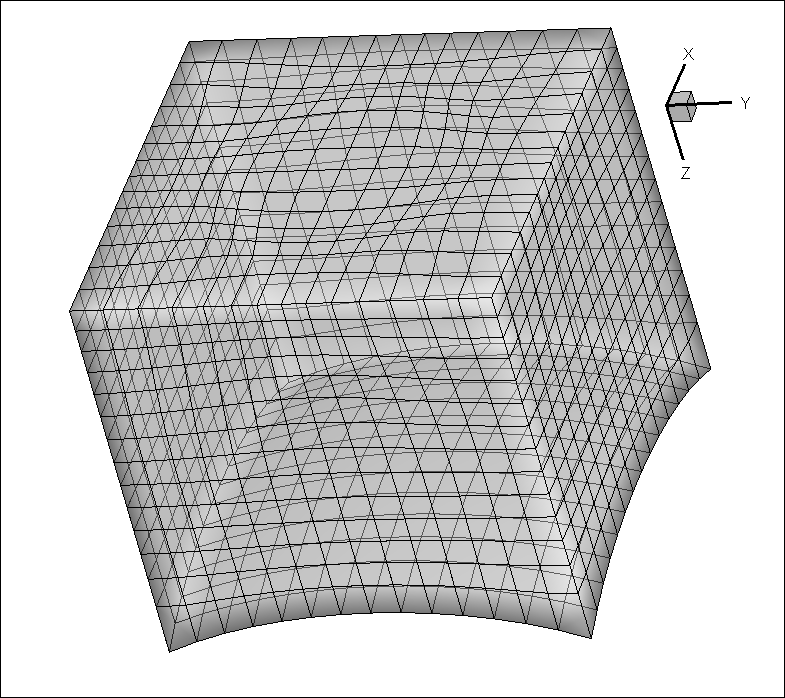}}
	\subfigure[t=1]{\includegraphics[width=8.5cm,height=7.5cm]{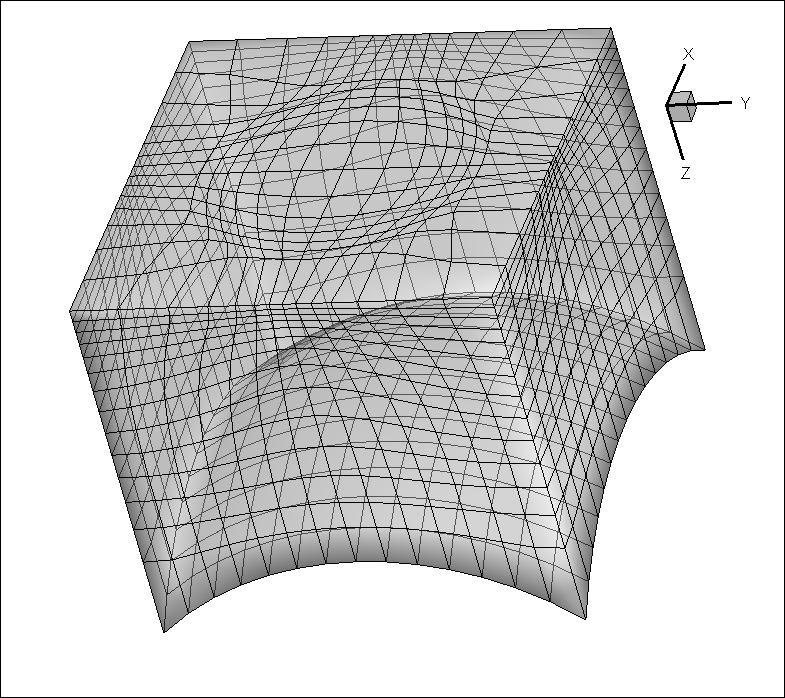}}
	\end{center}
\end{figure}
\begin{figure}[H]
	\caption{Deforming a tall brick to a horn}
	\begin{center}
	\subfigure[t=0]{\includegraphics[width=6cm,height=6.25cm]{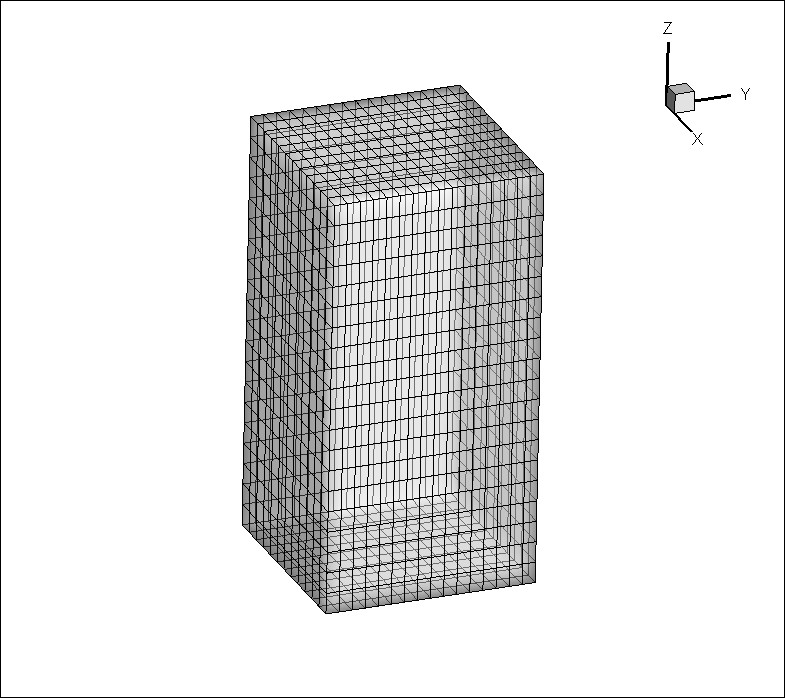}}
	\subfigure[t=0.5]{\includegraphics[width=6cm,height=6.25cm]{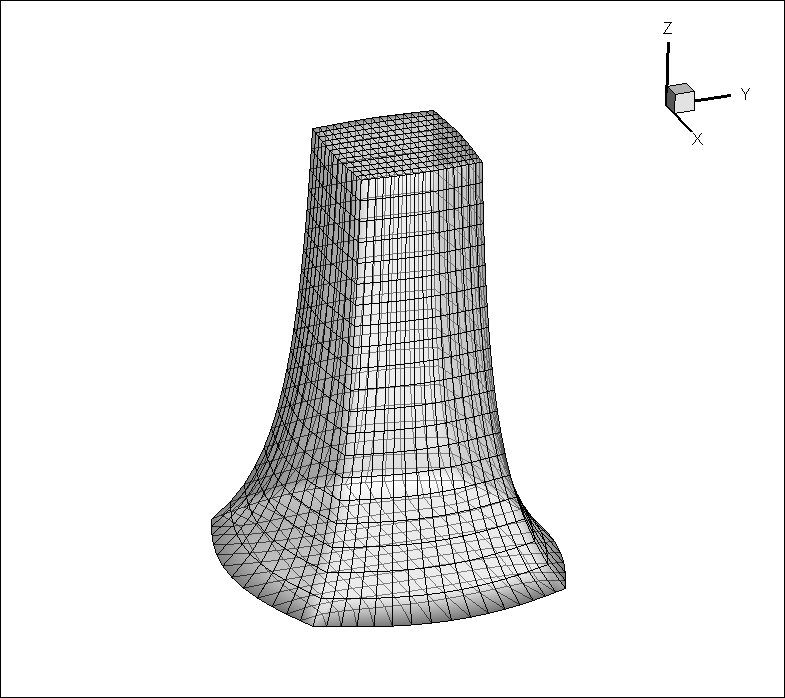}}
	\subfigure[t=1]{\includegraphics[width=8.5cm,height=9cm]{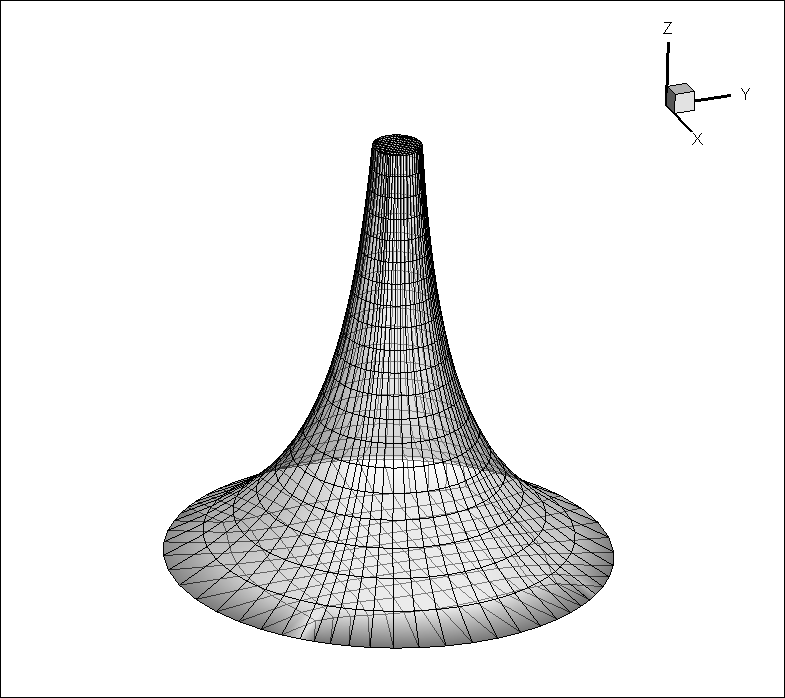}}
	\end{center}	
\end{figure}

\section{A New Algorithm for Higher Order Elements}
It is mentioned that the constraint $(\ref{eq:plm1})$ indicates the construction of a such diffeomorphism is started and deformed from $\pmb{\phi}(\pmb{\xi},0)=\pmb{id}(\pmb{\xi})$. In terms of mesh generation, the visualization of $\pmb{\phi}(\pmb{\xi},0)=\pmb{id}(\pmb{\xi})$ usually is represented by an uniform mesh. One may ask, what if the prescribed Jacobian determinant $f(\pmb{x},0)>0$ is not $1$, can we still construct such diffeomorphsim $\pmb{\phi}(\pmb{\xi},t)$ by the fashion of the $deformation$ $method$, i.e. can $\pmb{\phi}(\pmb{\xi},t)$ be found when $\pmb{\phi}(\pmb{\xi},t)\neq\pmb{id}(\pmb{\xi})$, which the visualization of it is not represented by an uniform mesh? The answer in general is $NO$. However, if there exists such diffeomorphism $\pmb{\varphi}_0(\pmb{\xi})$ describes the given domain with $J(\pmb{\varphi}_0(\pmb{\xi}))>0$ and satisfies (2.2), (2.3) and (2.4), then the answer is $YES$, which is also discussed in $\bf{Remark}$ $\bf{10}$ of \cite{XiChen}. 

The practical reason of understanding the $YES$ scenario is to assure our proposed refinement technique stands on a concrete theoretical ground. Because the diffeomorphism $\pmb{\phi}(\pmb{\xi},t)$ needs to be constructed from a domain not necessarily satisfies $\pmb{\phi}(\pmb{\xi},0)=\pmb{id}(\pmb{\xi})$, whose mesh representation is not necessarily uniform. Therefore, a more general statement of the problem goes as follows:

Let $\mathrm{\Omega}$ and $\mathrm{\Omega}_t \subset \mathbb R^n$, $n=2$, or $3$ and $0\leq{t}\leq{T}$, be a moving (includes fixed) domain and $\pmb{v}(\pmb{x},t)$ be the velocity field on $\partial\mathrm{\Omega_t}$, where $\pmb{v}(\pmb{x},t)\cdot{\pmb{\mathrm{n}}}=0$ on any part of $\partial\mathrm{\Omega}_t$ with slippery-wall boundary conditions and $\pmb{\mathrm{n}}$ is the outward normal vector of $\partial\mathrm{\Omega}_t$. Given diffeomorphism $\pmb{\varphi}_0:\mathrm{\Omega}\rightarrow\mathrm{\Omega}_0$ and scalar function $f(\pmb{x},t)>0 \in C^1(\pmb{x},t)$ on the domain $\mathrm{\Omega}_t \times [0,T]$, such that

	\begin{equation}\label{GP1}
	f(\pmb{x},0)=J(\pmb{\varphi}_0),
	\end{equation}
	\begin{equation}
	\int_{\mathrm{\Omega}_t} \dfrac{1}{f(\pmb{x},t)}d\pmb{x} = |\mathrm{\Omega}_0|.
	\end{equation}
A new (differ from the given $\pmb{\varphi}_0$) diffeomorphism 
	\begin{equation}
	\pmb{\phi}(\pmb{\xi},t):\mathrm{\Omega}_0\rightarrow\mathrm{\Omega}_t
	\end{equation}
such that $\forall t \in [0,T]$
	\begin{equation}\label{GP4}
	J(\pmb{\phi}(\pmb{\xi},t)) = f(\pmb{\phi}(\pmb{\xi},t),t)
	\end{equation}
can be constructed by solving the following differential equations, where $J(\pmb{\phi}(\pmb{\xi},t)) = \text{det}\bigtriangledown(\pmb{\phi}(\pmb{\xi},t))$ .

Similar to the previous case in section 2, we use this two-step procedure with appropriate modification to construct diffeomorphism $\pmb{\phi}(\pmb{\xi},t)$.
	\begin{itemize}
	\item First, determine $\pmb{u}(\pmb{x},t)$ on $\mathrm{\Omega}_t$ by solving
	\begin{equation}
	\left\{
		\begin{aligned}
		\text{div } \pmb{u}(\pmb{x},t)& = -\frac{\partial}{\partial t}(\dfrac{1}{f(\pmb{x},t)}) \\
		\text{curl } \pmb{u}(\pmb{x},t)& = 0\\
		\pmb{u}(\pmb{x},t)& = \dfrac{\pmb{v}(\pmb{x},t)}{f(\pmb{x},t)} \text{, on } \partial\mathrm{\Omega}_t
		\end{aligned}\right.
	\end{equation}
	\item Second, determine $\pmb{\phi}(\pmb{\xi},t)$ on $\mathrm{\Omega}_0$ by solving	
		\begin{equation}
		\left\{
			\begin{aligned}
			\frac{\partial \pmb{\phi}(\pmb{\xi},t)}{\partial t}& = f(\pmb{\phi}(\pmb{\xi},t),t) \pmb{u}(\pmb{\phi}(\pmb{\xi},t),t), \\
			\pmb{\phi}(\pmb{\xi},0)& = \pmb{\varphi}_{0}(\pmb{\xi})
			\end{aligned}\right.
		\end{equation}
	\end{itemize}

The above is the $deformation$ $method$ in a generalized version. The related theoretical derivations are exactly the same as in the case of $f(\pmb{x},0)=1$, in \cite{XiChen}.

Here, the refinement technique is to subdivide a given non-folding mesh (equivalent to have a given $\pmb{\varphi}_0$ s.t. $J(\pmb{\varphi}_0(\pmb{\xi}))>0$) or simply to the resulting coarse mesh determined from Algorithm 1, in order to get a intermediate mesh. Then, deform the intermediate mesh into a boundary conforming mesh. So, the construction of a new diffeomorphism $\pmb{\phi}(\pmb{\xi},t)$ is needed as the problem from $(\ref{GP1})$ to $(\ref{GP4})$ indicates. That leads to the idea of the following algorithm of local refinement. 

\begin{algorithm}
	\caption{$deformation$ $method$ w/ local refinements}
	\hrule % if wan to draw a line here, uncomment it
	\begin{itemize}
		\item Step 1: $\mathbf{Subdivision}$ of the given (coarse) mesh up to desired order $p$. 
		
		\item Step 2: Apply the corresponding Dirichlet b.c. to the new nodes and fix b.c. to the old nodes (or Neumann b.c. if it sit on the slippery part) of $\partial\mathrm{\Omega}_t$
		   
		\item Step 3: Apply $\mathbf{Algorithm}$ $\mathbf{1}$. ps: $dt$ can be chosen larger here, since the refined deformation happens only locally. 
	\end{itemize}
\end{algorithm}

It is noticed, the number of nodes of each subdivision is depend on the degree up desired order of HOE. Once the new nodes had been add, the intermediate mesh is formed. Before the construction of boundary conforming mesh, we need to apply the corresponding Dirichlet moving boundary conditions to the new nodes and Neumann boundary conditions to slippery-wall nodes on $\partial\mathrm{\Omega}_t$, because where they stand originally do not reflect the desired geometry. Additionally, we to apply Dirichlet fixed boundary conditions to old nodes from the coarse mesh since where they stand are reflecting the desired geometry. With appropriate applied corresponding boundary conditions, then we use Algorithm 1 which is based on the $deformation$ $method$ to the intermediate mesh to generate a finer boundary conforming mesh. 

In the next two examples show in Figure 8 and 9, in order to keep the coherence of our idea, we implemented the Algorithm 2 on the resulting coarse mesh with double sized $dt$ from Figure 2 and 3.

\begin{figure}[H]
	\caption{Continued from the resulting mesh of Figure 2, deform the new nodes from subdivision of example 1 to the same quarter disk}
	\begin{center}
	\subfigure[t=0]{\includegraphics[width=3.95cm,height=3.95cm]{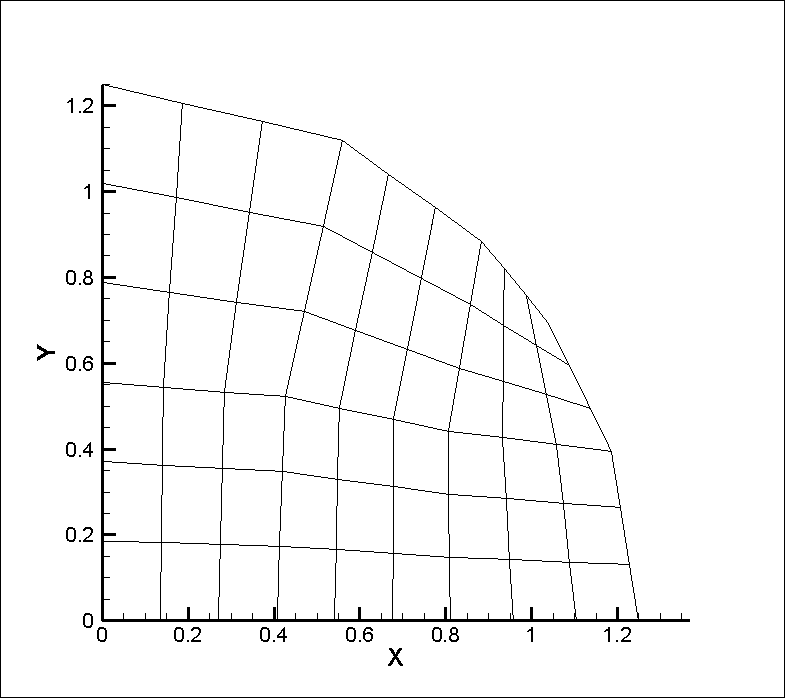}}
	\subfigure[t=0.6]{\includegraphics[width=3.95cm,height=3.95cm]{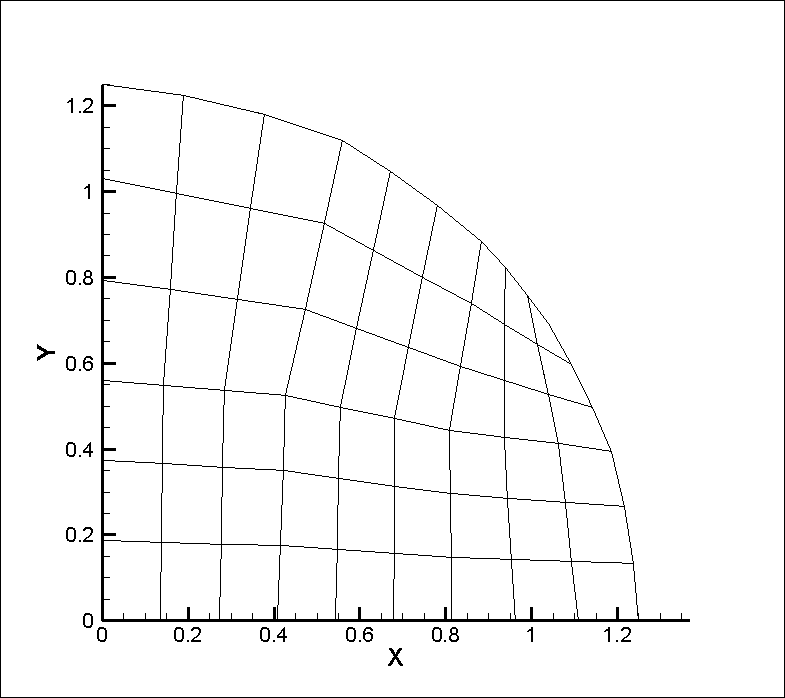}}
	\subfigure[t=1]{\includegraphics[width=3.95cm,height=3.95cm]{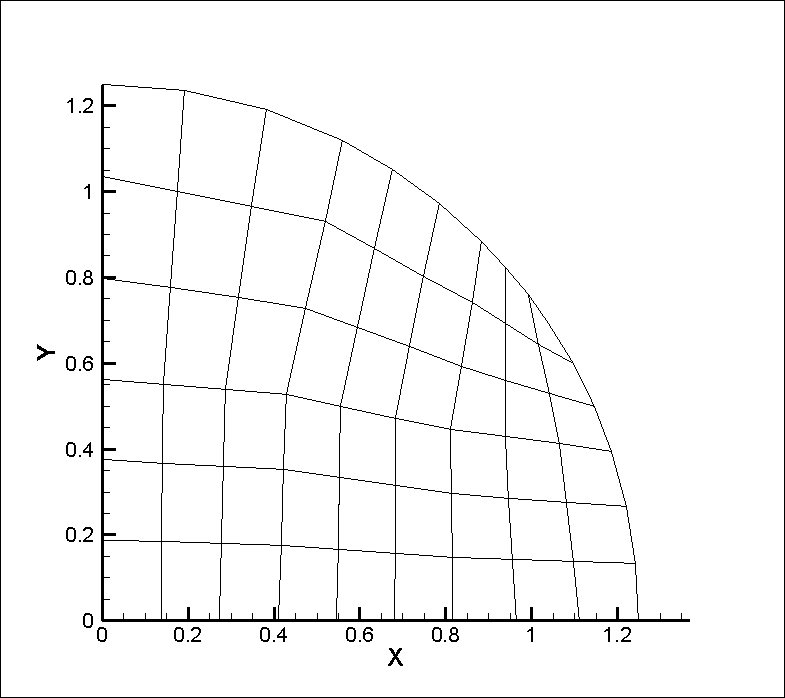}}
	\end{center}
	
\end{figure}

\begin{figure}[H]
	\caption{Deform the boundary nodes from the result of example 1 to the same quarter disk w/ $radius=1.25$}
	\begin{center}
	\subfigure[t=0]{\includegraphics[width=3.95cm,height=3.95cm]{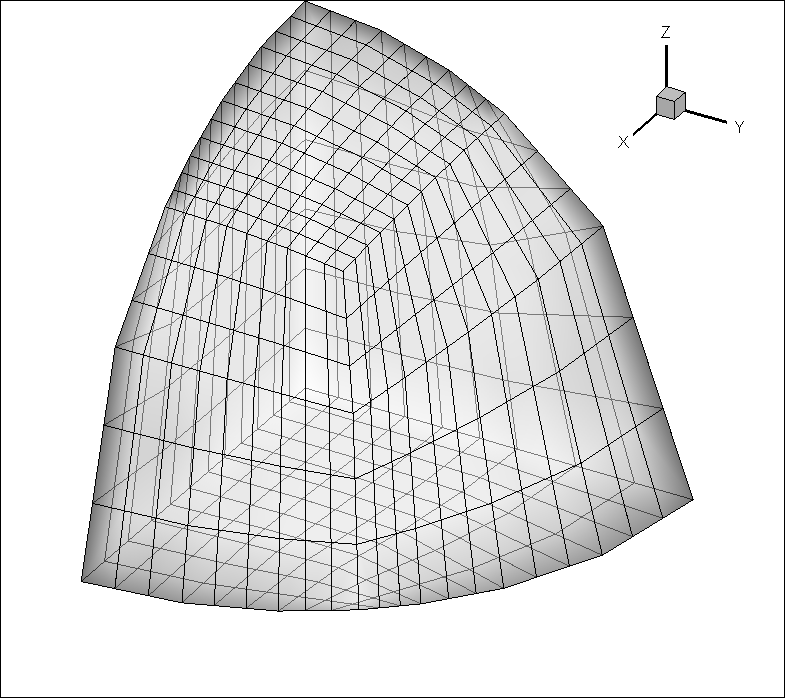}}
	\subfigure[t=0.6]{\includegraphics[width=3.95cm,height=3.95cm]{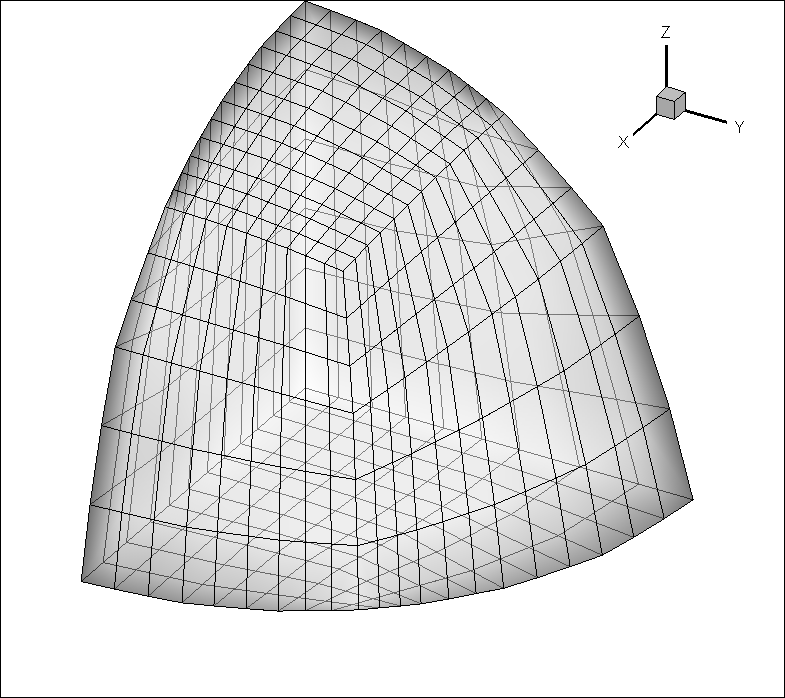}}
	\subfigure[t=1]{\includegraphics[width=3.95cm,height=3.95cm]{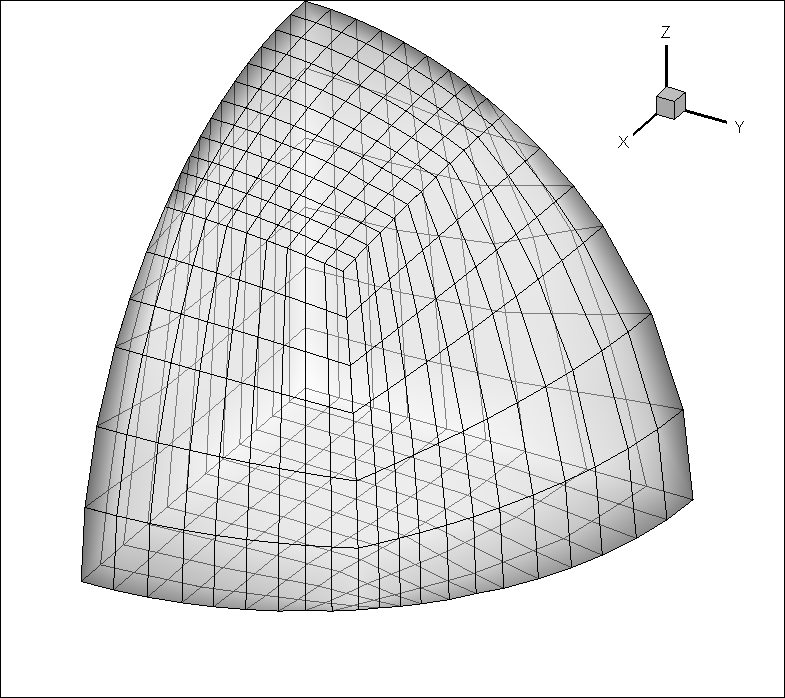}}
	\end{center}
	
\end{figure}

As the example showed above, the outcomes from the Algorithm 2, we are able to find more meaningful nodes in the sense of representing the original geometry of the coarse mesh intended to represent, i.e a boundary conforming mesh has achieved. That is what supports our motivation to have the realization the HOE mesh generation based on the $deformation$ $method$. The following algorithm is the proposed one by this paper. 
\begin{algorithm}[H]
	\caption{A new method for HOE mesh generation}
	\hrule % if wan to draw a line here, uncomment it
	\begin{itemize}
		\item Step 1: Apply $\mathbf{Algorithm}$ $\mathbf{1}$ to get a coarse mesh
		\item Step 2: Apply $\mathbf{Algorithm}$ $\mathbf{2}$ to the resulting coarse mesh to get a finer mesh		
		\item Step 3: $\mathbf{Interpolate}$ the desired geometry by appropriate choice of splines up to desired order $p$
		\item Step 4: $\mathbf{Plot}$ the interpolated data into the desired geometry based on the coarse mesh
	\end{itemize}
\end{algorithm}

We combined with appropriate interpolation techniques, for instance, polynomials with degree $p=3$, to define curves in 2D cases or surfaces in 3D case to represent the desired geometries. The simulates mesh may be different from the desired geometries, but with HOE, the higher order mesh is already better than the original coarse (linear) ones.

The improvement here is by using the $deformation$ $method$, we are able to located more nodes  that form a boundary conforming mesh for the desired geometry, which are necessary to the definition of HOE. Thus the HOEs can simply be defined by the cells of the higher order mesh. 

In the following HOE examples, the matlab interpolation package is used which computes with cubic-splines.
\begin{figure}[H]
	\caption{Mesh evolution example in 2D}
	\begin{center}
	\subfigure[coarse by Algorithm 1 ]{\includegraphics[width=3.95cm,height=3.95cm]{QuarterDisk4.png}}
	\subfigure[refined by Algorithm 2]{\includegraphics[width=3.95cm,height=3.95cm]{QuarterDisk_ref3.png}}
	\subfigure[HOE by Algorithm 3  ]{\includegraphics[width=3.95cm,height=3.95cm]{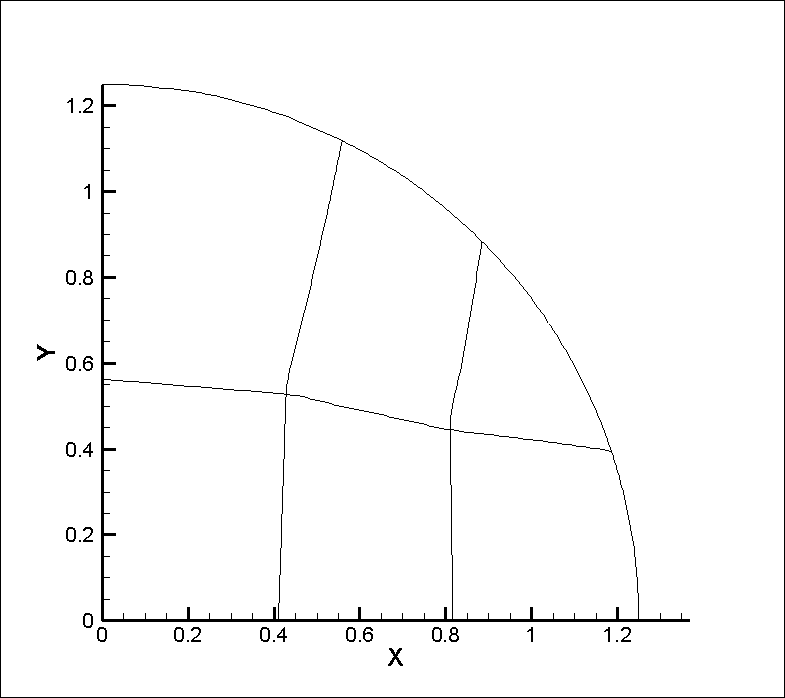}}
	\end{center}
\end{figure}

\begin{figure}[H]
	\caption{Mesh evolution example in 3D}
	\begin{center}
	\subfigure[coarse by Algorithm 1]{\includegraphics[width=3.95cm,height=3.95cm]{QuarterSemiSph10.png}}
	\subfigure[refined by Algorithm 2]{\includegraphics[width=3.95cm,height=3.95cm]{QuarterSemiSph10_ref10.png}}
	\subfigure[HOE by Algorithm 3]{\includegraphics[width=3.95cm,height=3.95cm]{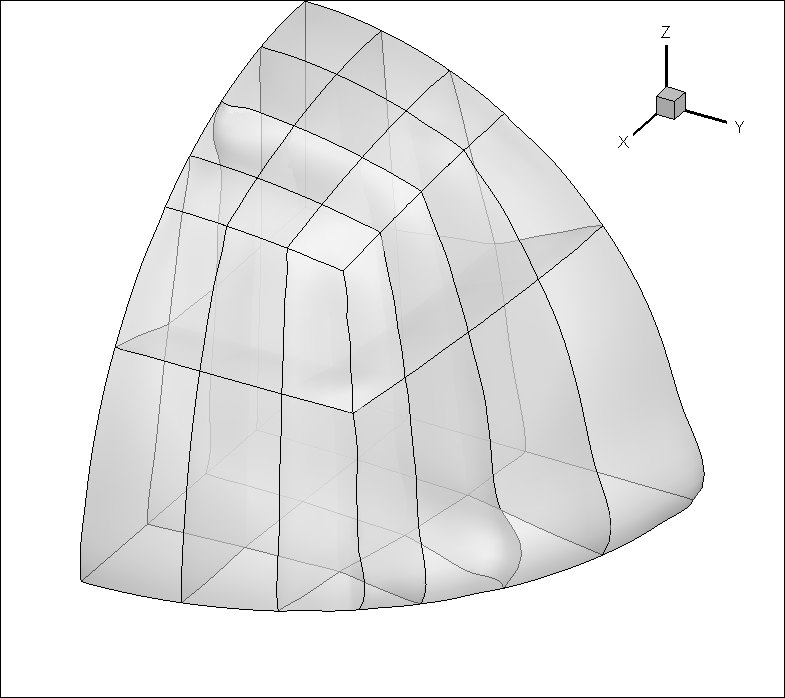}}
	\end{center}
\end{figure}

In fact, as Algorithm 3 states, it contains the procedures of Algorithm 1 and Algorithm 2, so the $deformation$ $method$ with LSFEM is not a trivial preliminary but is the core asset of our approach to higher order mesh generation. The following examples are more sophisticated ones that are achieved by Algorithm 3 which demonstrates flexibility and concision of our approach in acquiring higher order mesh for most common geometries. 

\begin{figure}[H]
	\caption{Deform a rectangle to water wave-like shape while emphasizing an interface of two different directed ellipsoids by Algorithm 1}
	\begin{center}
	\subfigure[t=0]{\includegraphics[width=5.25cm,height=5cm]{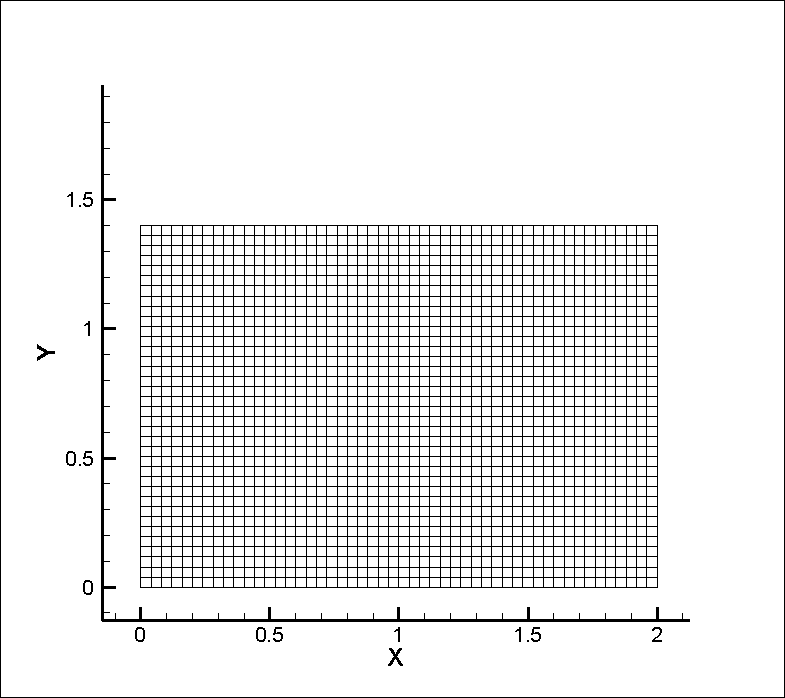}}
	\subfigure[t=0.3]{\includegraphics[width=5.25cm,height=5cm]{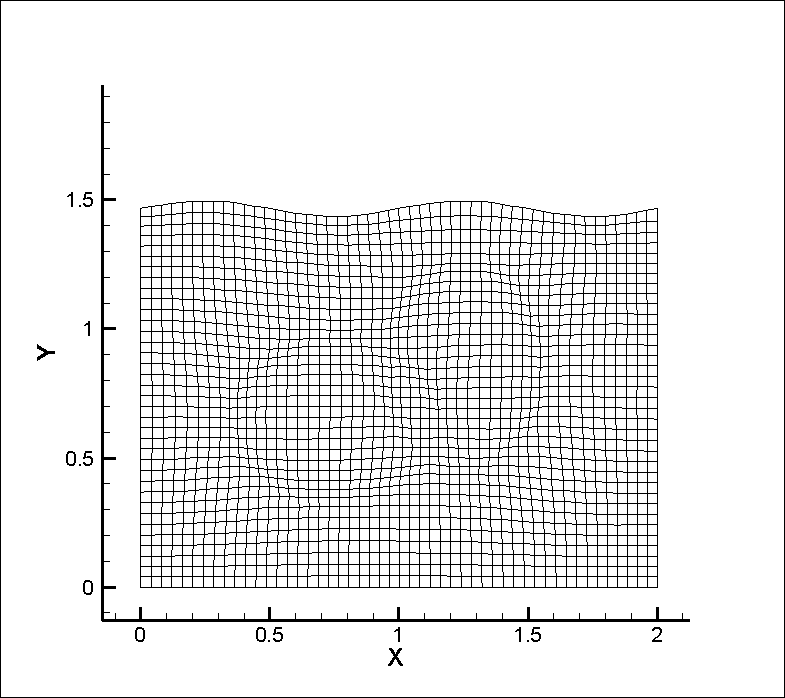}}
	\subfigure[t=0.5]{\includegraphics[width=5.25cm,height=5cm]{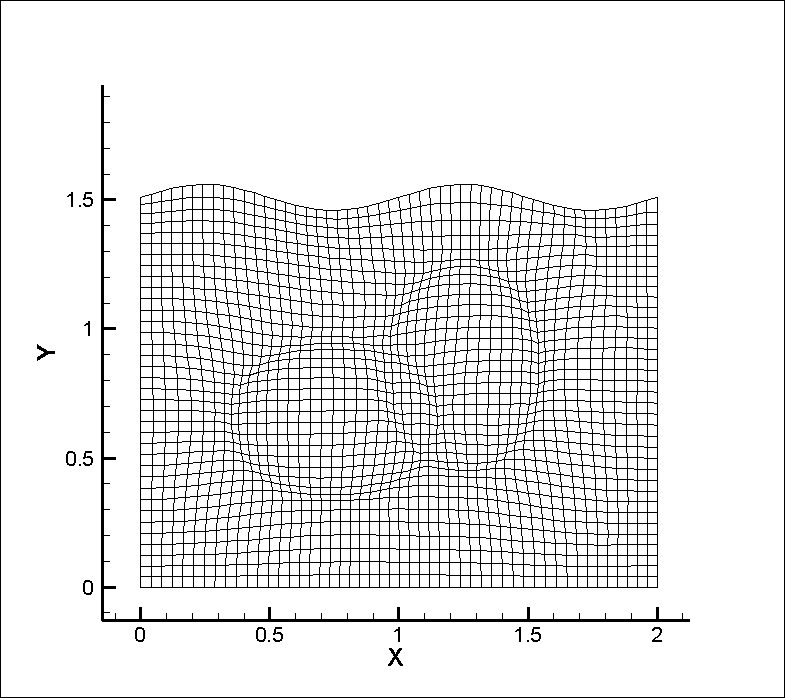}}
	\subfigure[t=0.8]{\includegraphics[width=5.25cm,height=5cm]{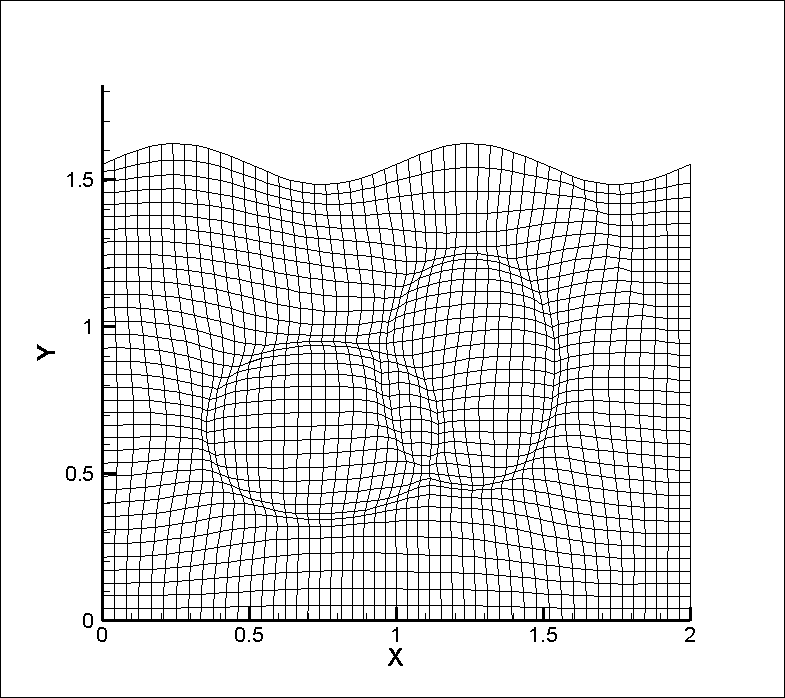}}
	\subfigure[t=1]{\includegraphics[width=7cm,height=6.5cm]{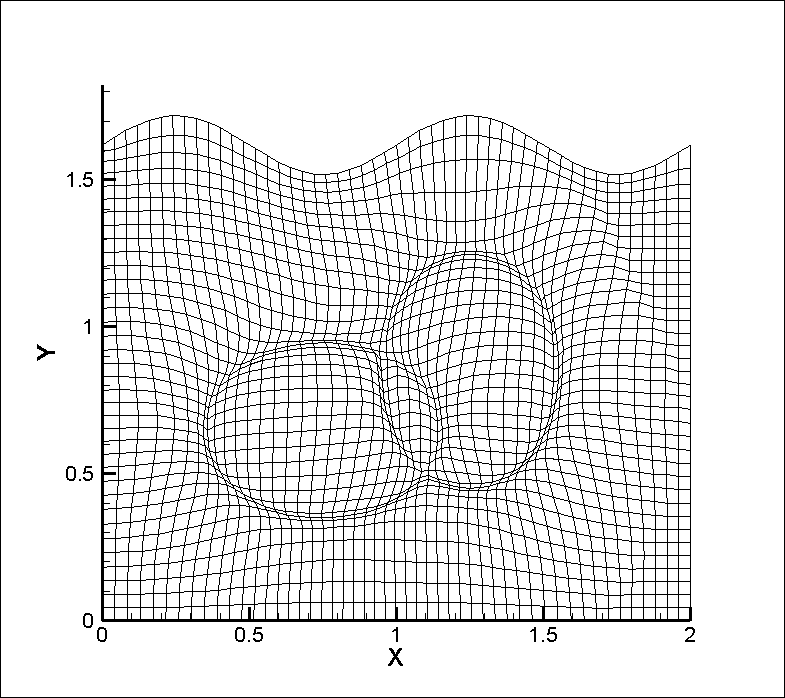}}
	\end{center}
\end{figure}
\begin{figure}[H]
	\caption{Deform the subdivided mesh to a finer one w/ Algorithm 2}
	\begin{center}
	\subfigure[t=0]{\includegraphics[width=5.25cm,height=5cm]{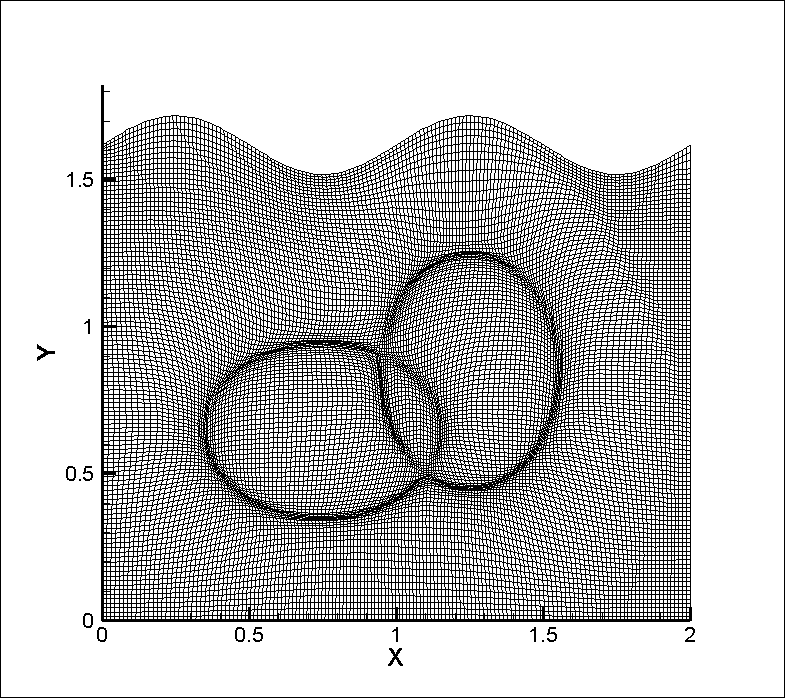}}
	\subfigure[t=0.3]{\includegraphics[width=5.25cm,height=5cm]{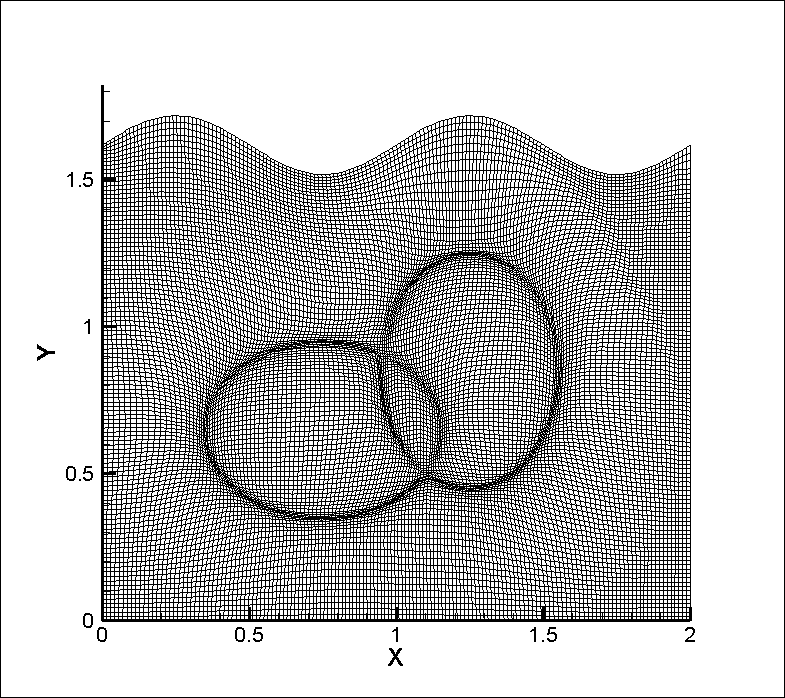}}
	\subfigure[t=0.8]{\includegraphics[width=5.25cm,height=5cm]{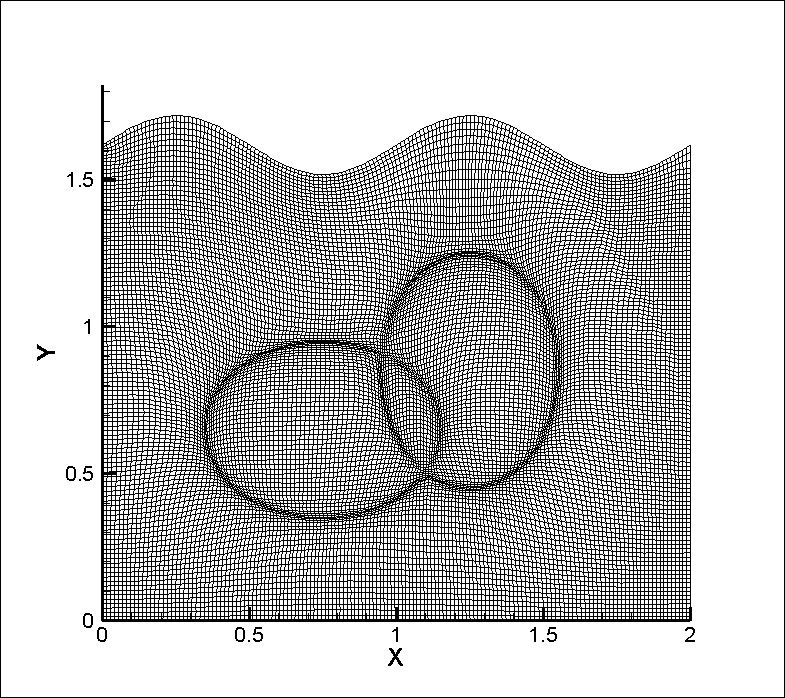}}
	\subfigure[t=1]{\includegraphics[width=7cm,height=6.5cm]{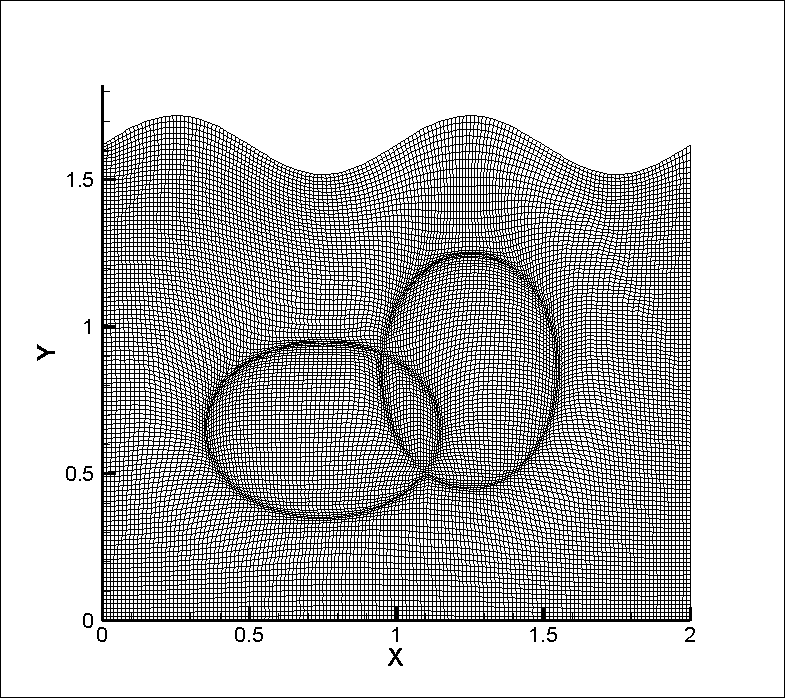}}
	\end{center}
\end{figure}
\begin{figure}[H]
	\caption{HOEs representation for the same mesh Figure 12 intend to represent}
	\begin{center}
	\subfigure[HOE by Algorithm 3]{\includegraphics[width=7cm,height=6.5cm]{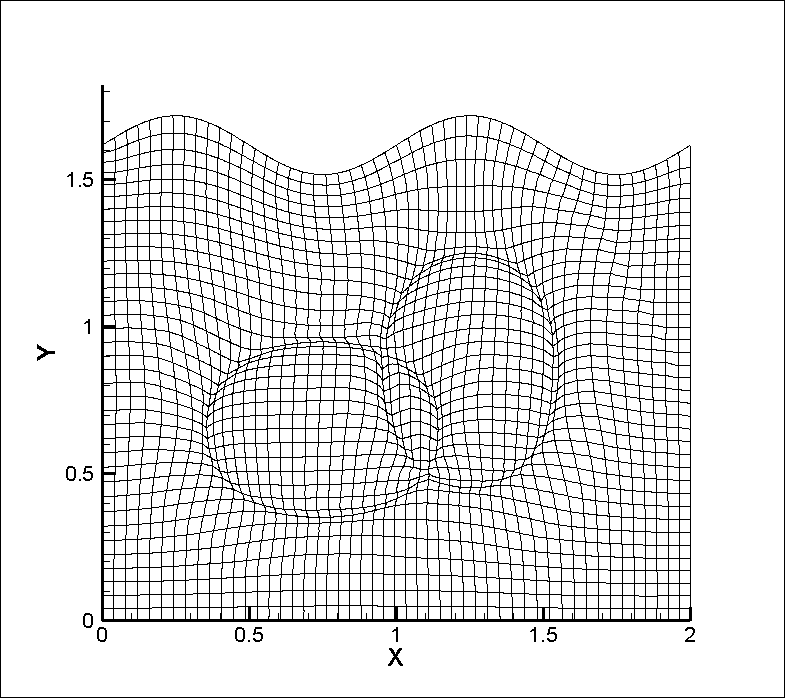}}
	\end{center}
\end{figure}
\begin{figure}[H]
 	\caption{Magnified views on the same example shown on Figure 12, 13 and 14 where the two emphasizing ellipses interface intersect each other}
 	\begin{center}
 	\subfigure[coarse by Algorithm 1]{\includegraphics[width=3.95cm,height=3.95cm]{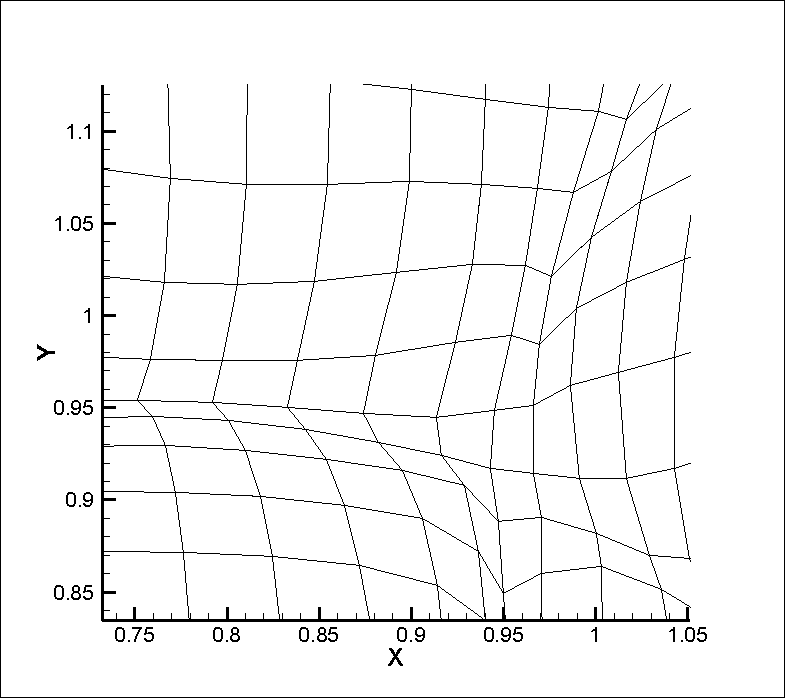}}
 	\subfigure[refined by Algorithm 2]{\includegraphics[width=3.95cm,height=3.95cm]{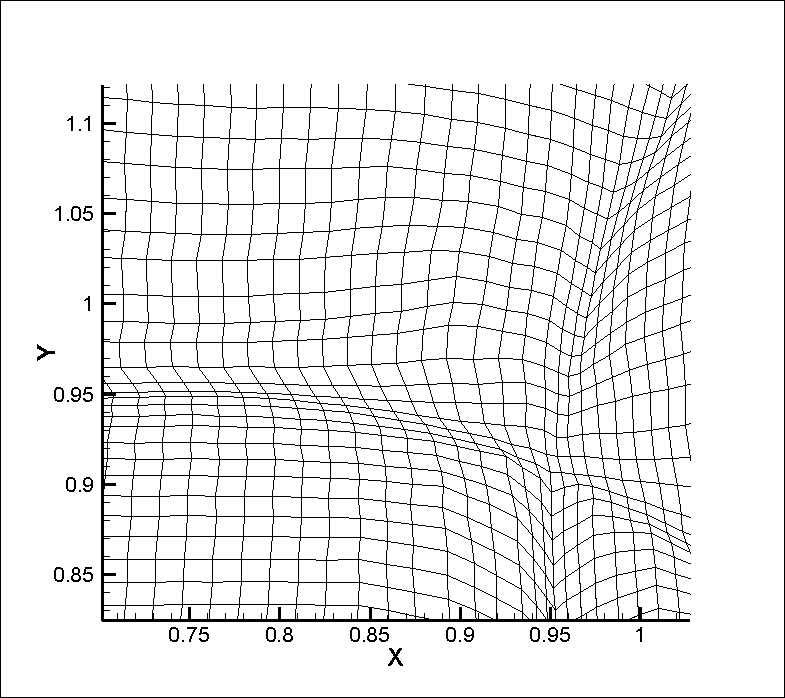}}
 	\subfigure[HOE by Algorithm 3]{\includegraphics[width=3.95cm,height=3.95cm]{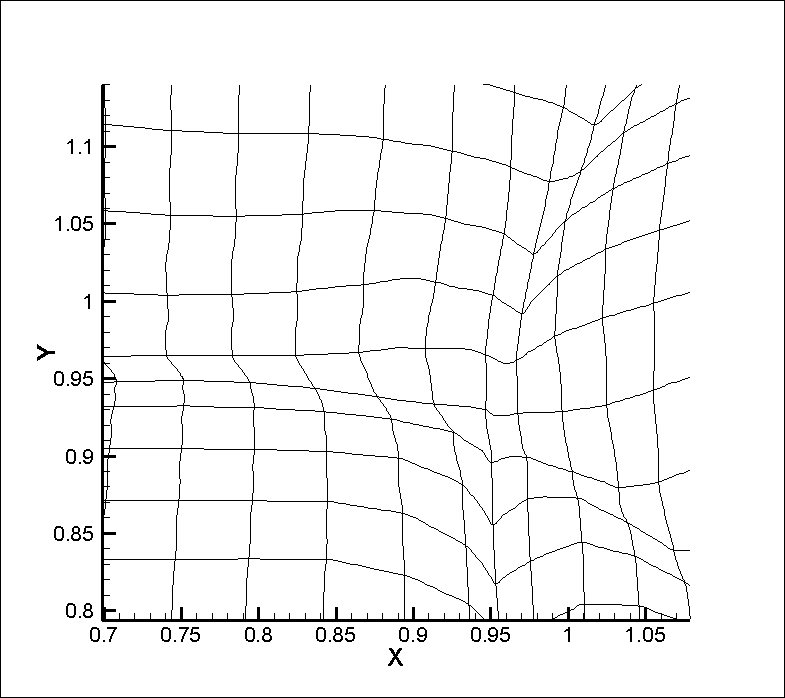}}
 	\end{center}
 \end{figure}
 
\begin{figure}[H]
	\caption{Deform a brick to an ellipsoid by Algorithm 1}
	\begin{center}
	\subfigure[t=0]{\includegraphics[width=5cm,height=4.65cm]{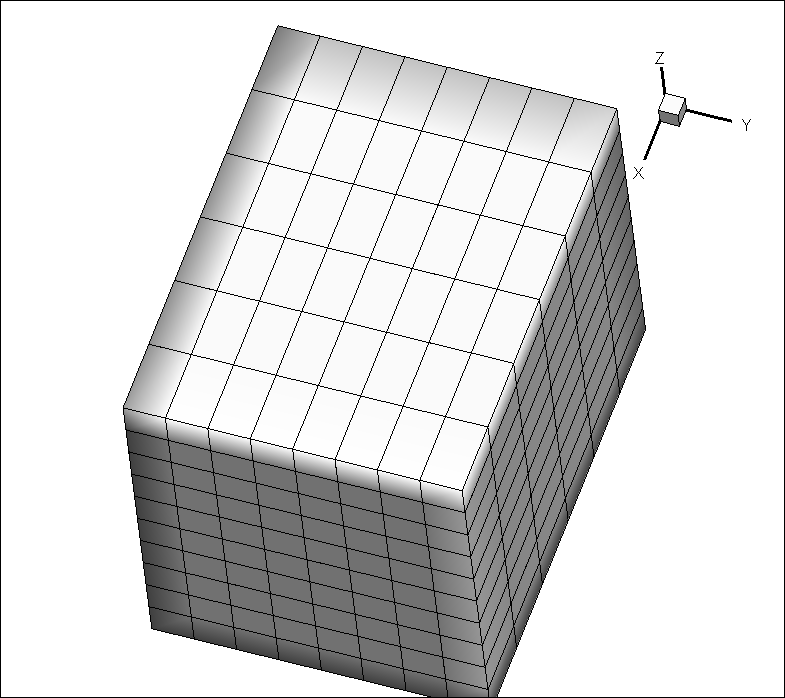}}
	\subfigure[t=0.3]{\includegraphics[width=5cm,height=4.65cm]{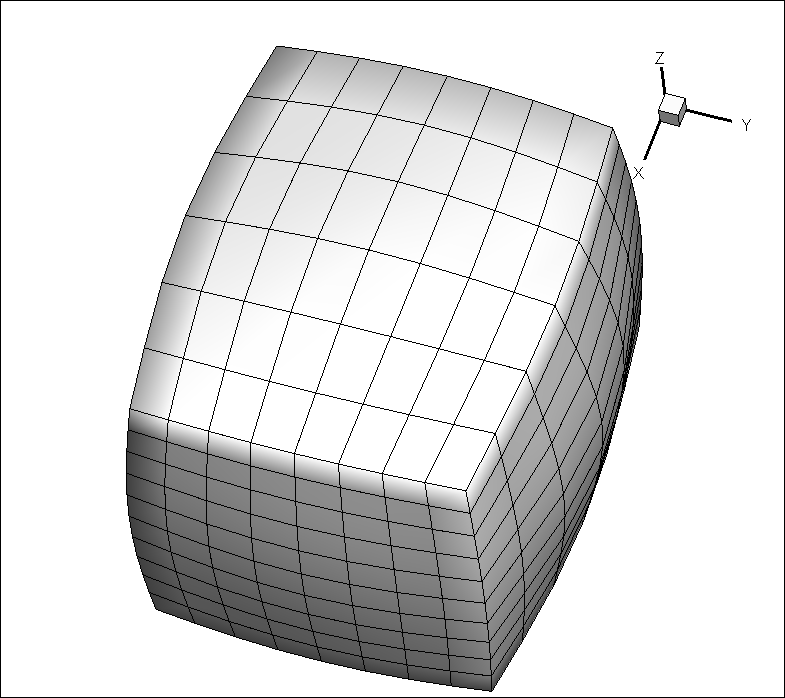}}
	\subfigure[t=0.5]{\includegraphics[width=5cm,height=4.65cm]{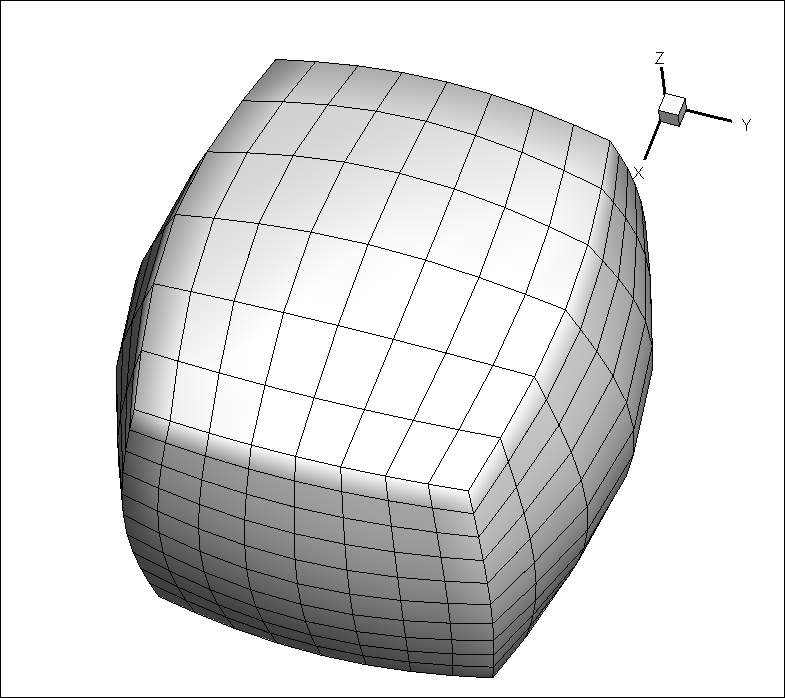}}
	\subfigure[t=0.8]{\includegraphics[width=5cm,height=4.65cm]{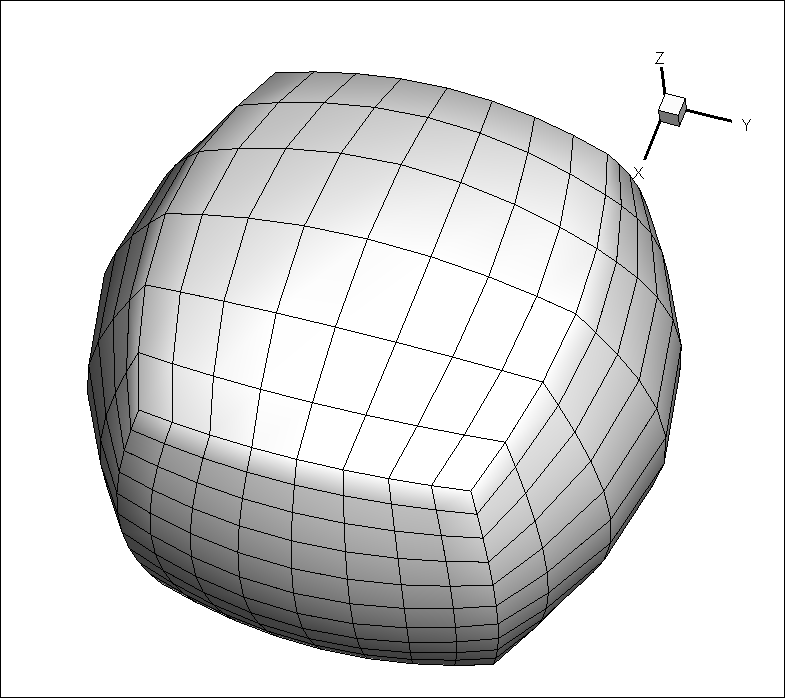}}
	\subfigure[t=1]{\includegraphics[width=7cm,height=6.5cm]{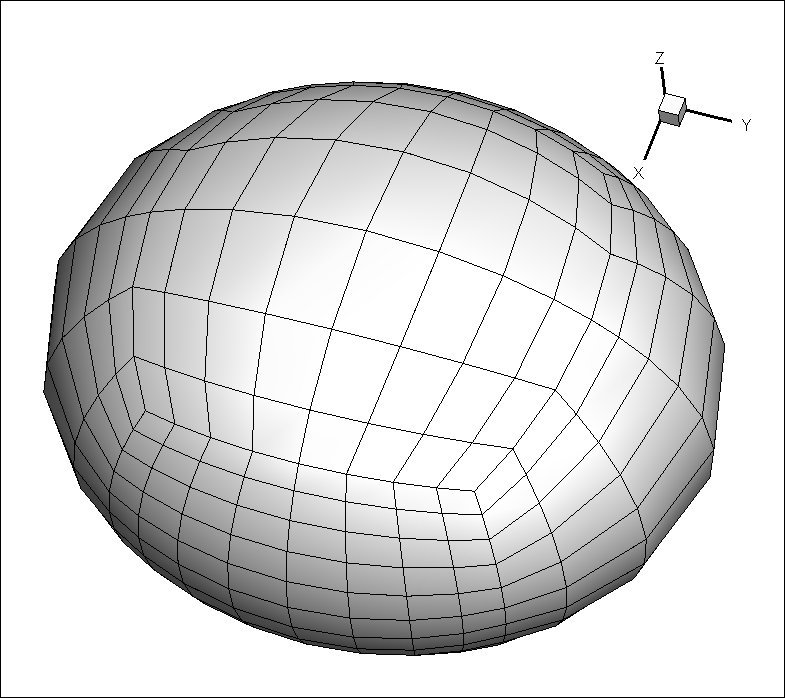}}
	\end{center}
\end{figure}

\begin{figure}[H]
	\caption{Deform the subdivided coarse ellipsoid to a finer one w/ Algorithm 2}
	\begin{center}
	\subfigure[t=0]{\includegraphics[width=5cm,height=4.25cm]{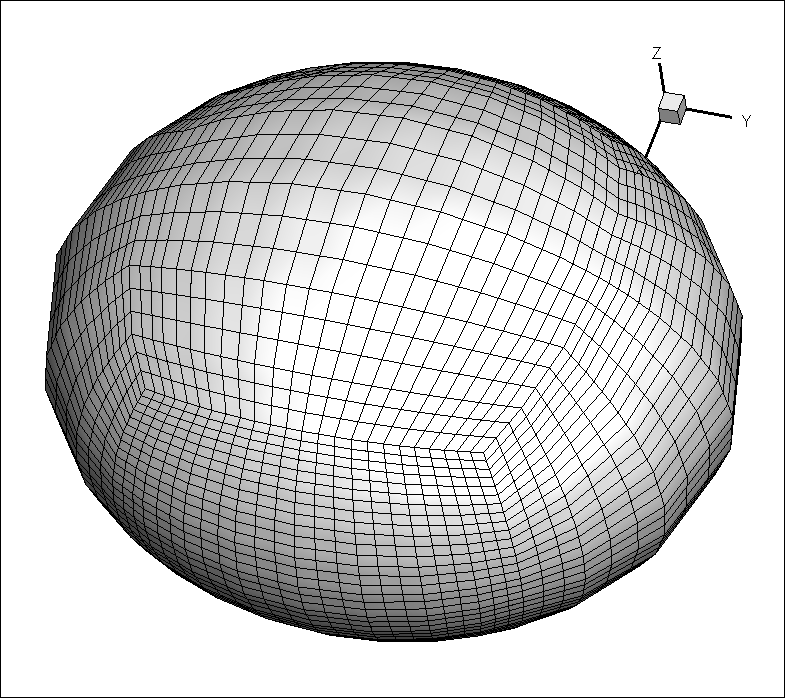}}
	\subfigure[t=0.3]{\includegraphics[width=5cm,height=4.25cm]{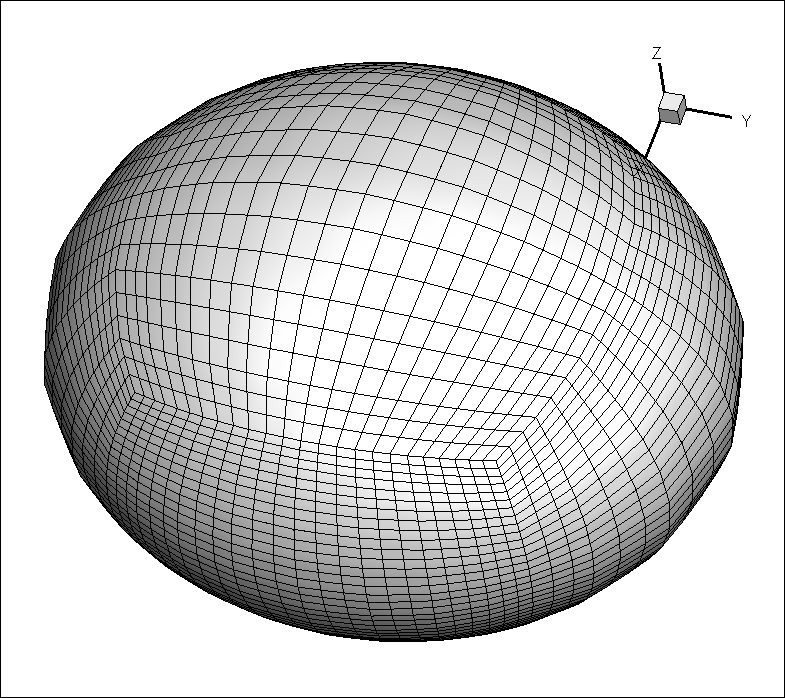}}
	\subfigure[t=0.8]{\includegraphics[width=5cm,height=4.25cm]{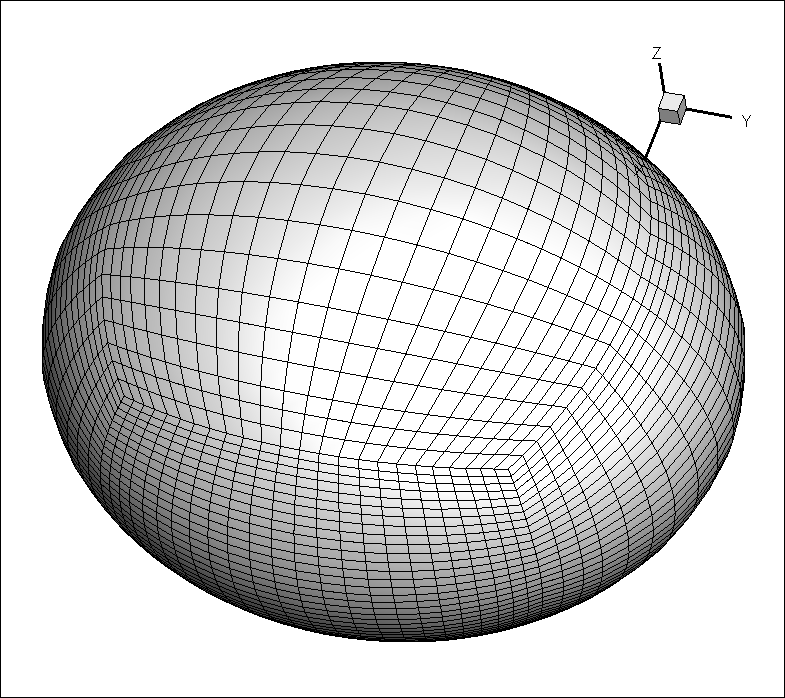}}
	\subfigure[t=1]{\includegraphics[width=7cm,height=6cm]{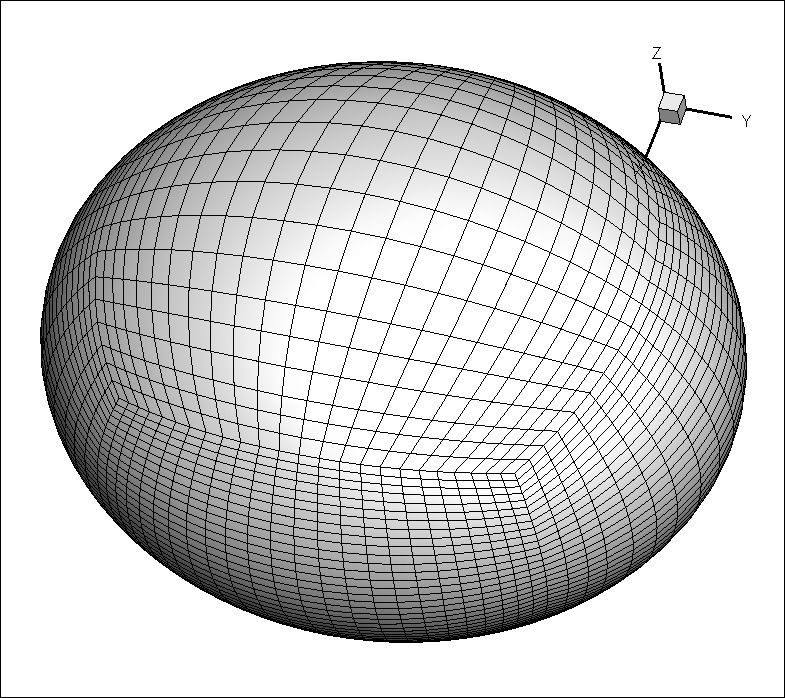}}
	\end{center}
\end{figure}

\begin{figure}[H]
	\caption{HOE representation for the same ellipsoid}
	\begin{center}
	\subfigure[HOE by Algorithm 3]{\includegraphics[width=7.5cm,height=6.4cm]{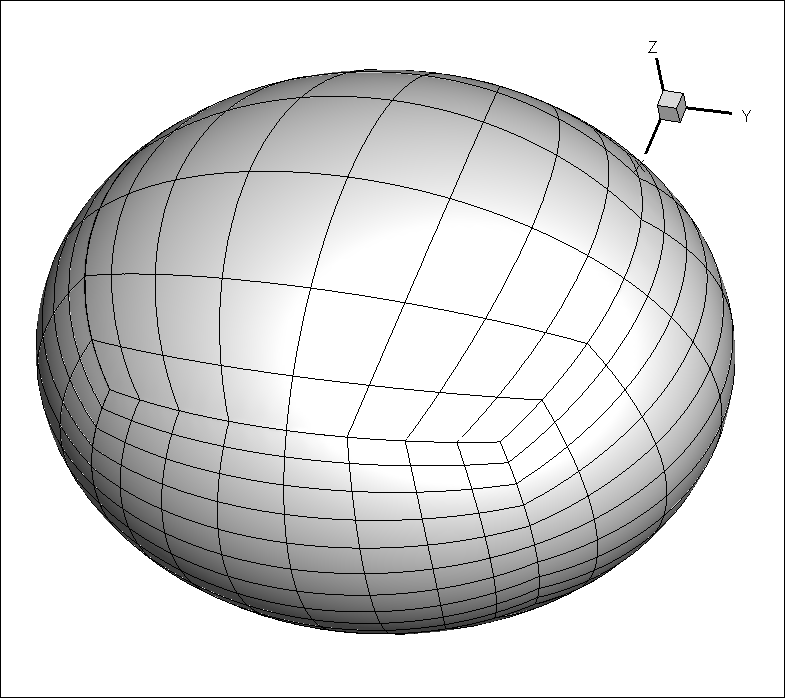}}
	\end{center}
\end{figure}

\section{Conclusions and Future work}

In this work, we formulated a new approach to higher order element generation. The main idea is to combine a local refinement technique with the $deformation$ $method$ on variable domains.

The new algorithm has a solid theoretical foundation. Moreover, its numerical implementation is based on solving a $\bf{divergence-curl}$ system by the Lease-Squares Finite Element Method. This numerical method is very well understood and efficient. 

The shown examples in this work are based on simple geometric models. Our goal is to develop a software package for real world problems in science and engineering.
%
% ---- Bibliography ----
%

\end{document}